\documentclass[fleqn,usenatbib]{mnras}
\usepackage{newtxtext,newtxmath}

\usepackage[T1]{fontenc}
\usepackage{ae,aecompl}
\usepackage{tabularx}

\DeclareRobustCommand{\VAN}[3]{#2}
\let\VANthebibliography\thebibliography
\def\thebibliography{\DeclareRobustCommand{\VAN}[3]{##3}\VANthebibliography}

\usepackage{graphicx}	
\usepackage{amsmath}	

\usepackage{amssymb}	
\usepackage{gensymb}

\newcommand{\msun}{M_\odot}

\newcommand{\kms}{{\rm km~s}$^{-1}$}

\newcommand{\Romeo}{\texttt{Romeo}}
\newcommand{\Juliet}{\texttt{Juliet}}

\newcommand{\tcoolsh}{t_{\rm cool}^{({\rm s})}}
\newcommand{\tff}{t_{\rm ff}}
\newcommand{\lcdm}{$\Lambda$CDM}
\defcitealias{Yu2021}{Y21}

\title[Born this way]{Born this way: thin disc, thick disc, and isotropic spheroid formation in FIRE-2 Milky-Way-mass galaxy simulations} \vspace{-0.1cm}

\author[Yu, Bullock, Gurvich et al.]{\hspace{-.01cm}Sijie Yu$^{1}$,  
James S. Bullock$^{1}$\thanks{E-mail: bullock@uci.edu},
Alexander B. Gurvich$^2$,
Zachary Hafen$^{1}$,
Jonathan Stern$^{3}$,\newauthor
Michael Boylan-Kolchin$^{4}$,
Claude-Andr\'e Faucher-Gigu\`ere$^2$,
Andrew Wetzel$^5$,
Philip F. Hopkins$^{6}$, \newauthor
Jorge Moreno$^7$
\vspace*{5pt} \\
\\
$^{1}$Department of Physics and Astronomy, University of California Irvine, CA 92697, USA \\
$^2$Department of Physics \& Astronomy and CIERA, Northwestern University, 1800 Sherman Ave, Evanston, IL 60201, USA \\
$^3$School of Physics \& Astronomy, Tel Aviv University, Tel Aviv 69978, Israel \\
$^4$Department of Astronomy, The University of Texas at Austin, 2515 Speedway, Stop C1400, Austin, TX 78712-1205, USA \\
$^5$ Department of Physics \& Astronomy, University of California, Davis, CA, USA 95616 \\
$^{6}$ TAPIR, Mailcode 350-17, California Institute of Technology, Pasadena, CA 91125, USA \\
$^7$ Department of Physics and Astronomy, Pomona College, Claremont, CA 91711, USA 
}

\date{Accepted XXX. Received YYY; in original form ZZZ}

\pubyear{2022}

\begin{document}
\label{firstpage}
\pagerange{\pageref{firstpage}--\pageref{lastpage}}
\maketitle

\begin{abstract}
We investigate the formation of  Milky-Way-mass galaxies using FIRE-2 \lcdm\ cosmological zoom-in simulations by studying the orbital evolution of stars formed in the main progenitor of the galaxy, from birth to the present day. We classify {\em in situ} stars as \textit{isotropic spheroid}, \textit{thick-disc}, and \textit{thin-disc} according to their orbital circularities and show that these components are assembled in a time-ordered sequence from early to late times, respectively.  All simulated galaxies experience an early phase of bursty star formation that transitions to a late-time steady phase.  This transition coincides with the time that the inner CGM virializes. During the early bursty phase, galaxies have irregular morphologies and new stars are born on radial orbits; these stars evolve into an isotropic spheroidal population today.  The bulk of thick-disc stars form at intermediate times, during a clumpy-disc ``spin-up'' phase, slightly later than the peak of spheroid formation.  At late times, once the CGM virializes and star formation ``cools down," stars are born on circular orbits within a narrow plane.  Those stars mostly inhabit thin discs today.  Broadly speaking, stars with disc-like or spheroid-like orbits today were born that way. Mergers onto discs and secular processes do affect kinematics in our simulations, but play only secondary roles in populating thick-disc and {\em in situ} spheroid populations at $z=0$. The age distributions of spheroid, thick disc, and thin disc populations scale self-similarly with the steady-phase transition time, which suggests that morphological age dating can be linked to the CGM virialization time in galaxies. 

\end{abstract}



\section{Introduction}

A long-standing quest in the field of galaxy formation is to understand why some stars are arranged in thinly-rotating discs while others inhabit more isotropic, spheroidal distributions. Remarkably, while easily formulated, a definitive answer to this question remains elusive.  

The Milky-Way (MW) mass scale is of particular interest: $M_\star \simeq 5 \times 10^{10}~\msun$ sits near an interesting crossover point, above which massive galaxies are typically early-type and spheroidal, and below which (central) galaxies become more disc-dominated and late-type  \citep{Bell03}.  Indeed, the MW-mass regime exhibits significant variance in morphological structure \citep{Freeman70,Kent85,Abraham03,Simard11,Bell17}.

It is possible to characterize the morphological components of galaxies in three broad components: thin discs, thick discs, and spheroids 
\citep{Oort1922,Lindblad1925,deVaucouleurs59,Tsikoudi79,Burstein79,YD06,vdKF2011}.Thin discs are systematically younger and more metal-rich than both spheroids and thick discs \citep{YD08,Comeron11}. By spheroid, we refer quasi-isotropic populations, which could include inner stellar halos \citep{Carollo07,Bonaca17} and isotropic bulge components. Note that bulge components in galaxies are usually classified in one of two ways: flattened,  mildly rotating \textit{pseudo bulges}  or more spheroidal \textit{classical bulges} \citep{Kormendy04,Gao20}.  
Most stars in classical bulges appear to be formed at high redshift, with only minor growth at late times \citep{Renzini99}.  

Among the most important ideas to emerge in galaxy formation theory is that rotationally-supported disc galaxies form naturally as a result of angular momentum conservation, with the angular momentum source driven by gravitational collapse in an expanding universe \citep{Peebles69,Fall79}.  Rotating galaxies may then be seen as a natural starting point for subsequent morphological evolution  \citep{Hodge19,Rizzo20,Tamfal21,Kretschmer21}.
In the simplest version of this scenario, thin discs form early and continuously, with thick discs and bulges arising only from secular evolution and/or mergers.
A more nuanced version of this idea, motivated by \lcdm\ galaxy formation simulations, is that the first discs to emerge in the early universe are thick and turbulent, with thin-disc galaxies developing only late in cosmic assembly \citep{Brook12,Wuyts12,Bird13,Park19,Bird20,Yu2021,Alvaro22}. 

Observationally, disc galaxies at higher redshift do appear to be more disordered and clumpy \citep[]{Elmegreen07,Shapiro2008,Genzel2008,Overzier2010,Elmegreen17,Osborne2020} and only later does star formation begin to occur primarily in extended thin discs \citep{Kassin12}. This behavior is consistent with a picture where the discs evolve in quasi-stable equilibrium, with higher ISM velocity dispersion at early times owing to an increasing gas fraction \citep{FG13,Wisnioski15,Ceverino17,Bird20}.  Interestingly, some of the first observations with {\em JWST} suggest that discs may be more common in the early universe than previously believed \citep[][]{Robertson22,Ferreira22}.

The rotational and structural properties of pseudobulges are indicative of a formation channel linked closely to disc formation and/or disc/bar evolution  \citep{Okamoto13,Kormendy15,Spiegel19,Devergne20}.  Classical bulges, on the other hand, appear to be less connected to disc properties and it remains unclear if their formation is linked to or completely disjoint from disc assembly.  Traditional scenarios posit that classical bulges emerge from the mergers of discs \citep{vdB98,Hopkins10,Bois10,Kannan2015}, though there is theoretical and observational evidence that bulges can form in multiple ways \citep[e.g.][]{Obreja13,Seidel15}.  Interestingly, direct measurements suggest that bulge mass is not linked closely to galaxy merger history at the Milky Way scale \citep{Bell17}. Secular process, like rapid gas inflow to the galaxy center \citep{Scannapieco09}, or the migration of giant gaseous clumps \citep{Minchev10,Ceverino15}, could naturally produce quasi-isotropic bulge formation.

Similarly, the inner stellar halos of galaxies may be populated by multiple channels.
Today there is general agreement that  accretion is responsible for the majority of stars in the  {\em outer} stellar halos of galaxies \citep{Bullock2001,Bullock&Johnston05,DeLucia08,Bell2008,McConnachie2009}, though a fraction could originate in outflows from the main galaxy \citep{Yu20}. 
Inner stellar halo, on the other hand, is likely populated by both accreted stars \citep[e.g.][]{Helmi_2018,Mackereth_2018,Simon2019} and stars that were born 
 within the inner galaxy \citep{Carollo07,Cooper15}.  This {\it in situ} inner stellar halo is often believed to consist of stars that were born on orbits initially confined them to a disc but became heated to eccentric orbits by mergers \citep{Zolotov2009,Purcell10} or by potential fluctuations from explosive feedback events \citep{El-Badry18}.

Alternatively, inner halo or spheroid formation  could arise from an earlier, very turbulent phase of galaxy assembly with no need for coherent rotation. Interestingly, this possibility bares considerable resemblance to a conjecture put forth by \citet{Larson1976}, who used numerical experiments to suggest that spheroidal/bulge components arise in an early stage of rapid star formation, while the formation of  discs requires a later stage of slower star formation to allow the residual gas to settle into a disc before forming stars \citep[see also][]{Gott76,ELS}.  Larson concluded that ``Spheroidal systems may thus form from gas which experiences strong turbulence or cloud collisions, and disc systems may form from more quiescent residual gas in which collisions are less important.''

The Milky Way provides a detailed touchstone for testing these ideas.  Our galaxy has both thin and thick disc components \citep{GR83,juric08,Bensby11,BovyRix13,Hayden15}.  Its inner  stellar halo appears to be populated by both {\em in situ} and accreted components \citep{Carollo07,Helmi_2018,Belokurov18}.  The Galaxy has a small classical bulge \citep{Kunder16} and a more dominant pseudobulge \citep{Gonzalez16}, which appears to resemble the thick disc in chemical properties and formation time \citep{AB10,Haywood18,DiMatteo19}. 

Recently, chemo-dynamical data sets have started to uncover exciting clues to the origin of the rotating components of the Galactic disc.  
\cite{BK22} used APOGEE and \textit{Gaia} data to identify a characteristic ``spin up'' metallicity for Milky Way stars.  Specifically,
{\em in-situ} stars show a rapid increase in net rotation as a function of metallicity at $[{\rm Fe/H}] \simeq -1$, from a median tangential velocity of $\sim 0$ \kms\ (typical of a spheroid) to $\sim 100$ \kms\ (typical of a {\em thick} disc). 
This feature may point to transition epoch from  disordered kinematics to increasingly coherent rotation. Similarly, \citet{Conroy22} have used H3 Survey spectroscopy and \textit{Gaia} astrometry to identify a transition time where star formation efficiency rapidly increased while simultaneously stellar kinematics become more disc-like. Stars that formed before this time retain an isotropic velocity distribution. There are features of this picture, where there is an early ``spin-up" phase that follows a less well-ordered phase, that are quite similar to those seen in cosmological simulations of disc galaxy formation \citep[e.g.][]{Park21}.

In what follows, we use FIRE-2 \citep{Hopkins17} simulations to study the orbital properties of stars formed in the main progenitors of Milky Way size galaxies, from birth to the present day, in order to gain insight into the origin of thin discs, thick discs, and {\em in situ} isotropic spheroids. It extends work from a series of FIRE-2 papers that have examined the relationship between star formation burstiness, galaxy kinematics, galaxy morphology, metallicity gradients, and the development of a hot gaseous halos around galaxies \citep{Ma2017,Stern20,Yu2021,Bellardini22,Gurvich2022,Hafen2022}. In particular, these simulations have revealed a correlation between internal galaxy properties and the mode of gas deposition from the circum-galactic medium (CGM) into the interstellar medium (ISM) \citep{Stern20}.  As galaxy halos evolve from low mass to high, the inner CGM virialization, star formation transitions from ``bursty'' to ``steady,'' and stellar-driven galaxy-scale outflows are suppressed. The bulk of thick-disc stars form  prior to this transition, and this gives rise to a tight correlation between the ages of thick-disc stars and the end of the bursty phase \citep{Yu2021}.  CGM virialization also drives an abrupt change in the angular momentum coherence of accreting gas \citep{Hafen2022}.  Only after this time do stars form along a single long-lived plane in circular orbits, making possible the formation of a  {\em thin} disc \citep{Yu2021,Hafen2022}.  Conversely, during the earliest epochs, the ISM has a quasi-spheroidal morphology, and negligible rotation support \citep{Gurvich2022}. Stars formed during this phase may naturally produce a population of centrally-concentrated old stars that are isotropic and spheroidal in nature, with qualitative resemblance to stars that may contribute to classical bulges and/or inner stellar halos today.

In Section 2 we provide an overview of our simulations and the kinematic definitions (thin, thick, spheroid) we adopt in our analysis. In Section 3 we present results on the dynamical evolution of galaxy populations with time.  Section 4 is reserved for discussion and conclusions.

\begin{table*}
  \caption{The five simulations we employ for the bulk of this work are summarized in the top section of this table. The second set of seven are used only in Section \ref{sec:sample_trend}.  We list the following: the name of the zoom-in target halo, the stellar mass ($M_{\star}$) within the central 20 kpc of the halo at $z=0$, the radius (${R_\mathrm{90}}$) enclosing 90\% of $M_{\star}$, the halo virial mass using the \citet{Bryan98} definition ($M_{\mathrm{halo}}$), the halo virial radius ($R_{\mathrm{halo}}$), the resolution of each simulation quantified by the initial baryonic particle mass ($m_{\mathrm{i}}$), and the reference that first introduced each halo at the quoted targeted resolution. The remaining columns present derived quantities: the lookback time to the end of the bursty phase/onset of the steady phase ($t_{\rm B}$), the mass-weighted thin-disc fraction ($f_{\rm thin \, disc \, m}$), and the luminosity-weighted thin-disc fraction ($f_{\rm thin \, disc \, l}$). Hosts with names starting with `m12' are isolated configurations selected from the Latte suite, whilst the rest are in LG-like pairs from the ELVIS on FIRE suite. The four galaxies marked with an asterisk correspond to minor mergers taking place after the onset of the steady phase. 
  The haloes in each list are ordered by decreasing $t_{\rm B}$. } 
	\centering 
	\label{tab:info}
	\begin{tabularx}{\textwidth}{Xccccccccc}
		\hline
		\hline  
		Simulation & $M_{\star}$ & ${R_\mathrm{90}}$ &  $M_{\mathrm{halo}}$ & $R_{\mathrm{halo}}$ & $m_{\mathrm{i}}$ & $t_{\mathrm{B}}$ & $f_{\mathrm{thin\ disc\ m}}$ & $f_{\mathrm{thin\ disc\ l}}$ & Reference \\
		Name  &  $[M_{\odot}]$ & [kpc] & $[M_{\odot}]$ & [kpc] & $[M_{\odot}]$ & [Gyr] & ($M$ weighted) & ($L$ weighted) \\
		\hline\\[-0.32cm]
		\texttt{Romeo} & 7.4$\times$10$^{10}$ & 13.3 & 1.0$\times$10$^{12}$ & 317 & 3500 & 6.52 & 0.45 & 0.70 & A\\
		\texttt{m12b}* & 8.1$\times$10$^{10}$ & 9.8 & 1.1$\times$10$^{12}$ & 335 & 7070 & 6.32 & 0.37 & 0.64 & A\\
		\texttt{m12i} & 6.1$\times$10$^{10}$ & 12.8 & 9.2$\times$10$^{11}$ & 318 & 7070 & 5.11 & 0.33 & 0.64 & C\\
		\texttt{m12f}* & 8.6$\times$10$^{10}$ & 11.0 & 1.3$\times$10$^{12}$ & 357 & 7070 & 4.36 & 0.33 & 0.62 & B\\
		\texttt{Juliet} & 4.2$\times$10$^{10}$ & 9.6 & 8.5$\times$10$^{11}$ & 302 & 3500 & 4.40 & 0.30 & 0.62 & A\\
		\hline\\[-0.31cm]
		\texttt{Remus} & 5.1$\times$10$^{10}$ & 12.3 & 9.7$\times$10$^{11}$ & 320 & 4000 & 5.88 &0.36 & 0.62  & D\\
		\texttt{Louise} & 2.9$\times$10$^{10}$ & 12.0 & 8.5$\times$10$^{11}$ & 310 & 4000 & 5.56 & 0.32 & 0.65  & A\\
		\texttt{Romulus} & 1.0$\times$10$^{11}$ & 14.2 & 1.5$\times$10$^{12}$ & 375 & 4000 & 4.90 & 0.37 & 0.69  & D\\
		\texttt{m12m} & 1.1$\times$10$^{11}$ & 11.3 & 1.2$\times$10$^{12}$ & 342 & 7070 & 3.81 & 0.34 & 0.58 & E\\
		\texttt{m12c}* & 6.0$\times$10$^{10}$ & 9.7 & 1.1$\times$10$^{12}$ & 328 & 7070 & 3.70 & 0.32 & 0.62  & A\\
		\texttt{Thelma}* & 7.9$\times$10$^{10}$ & 12.4 & 1.1$\times$10$^{12}$ & 332 & 4000 & 2.57 & 0.27 & 0.57  & A\\
        \texttt{m12w} & 5.8$\times$10$^{10}$ & 8.7 & 8.3$\times$10$^{11}$ & 301 & 7070 & 0.0 & 0.24 & 0.43 & F\\
		\hline\\[-0.32cm]
	\end{tabularx}
	\raggedright
	\textit{Note}: The references are: 
	A:~\cite{Garrison-Kimmel19},
	B:~\cite{Garrison-Kimmel17}, 
	C:~\cite{Wetzel16}.
	D:~\cite{Garrison-Kimmel19_2}
	E:~\cite{Hopkins17}, 
	F:~\cite{Samuel20}.
	\label{tab:one}
\end{table*}

\section{Simulations and methods}
\label{sec:sims}

\subsection{FIRE-2 simulations of Milky-Way-mass galaxies}
\label{sec:sims_intro}

Our analysis utilises cosmological zoom-in simulations performed with the multi-method gravity plus hydrodynamics code {\small GIZMO} \citep{Hopkins15} from the Feedback In Realistic Environments (FIRE) project\footnote{\url{https://fire.northwestern.edu/}}. We rely on the FIRE-2 feedback implementation \citep{Hopkins17} and the mesh-free Lagrangian Godunov (MFM) method. The MFM approach provides adaptive spatial resolution and maintains conservation of mass, energy, and momentum. FIRE-2 includes radiative heating and cooling for gas across a temperature range of $10-10^{10}$~K. Heating sources include an ionising background \citep{Faucher2009}, stellar feedback from OB stars, AGB mass-loss, type Ia and type II supernovae, photoelectric heating, and radiation pressure, with inputs taken directly from stellar evolution models. The simulations self-consistently generate and track 11 elemental abundances (H, He, C, N, O, Ne, Mg, Si, S, Ca, and Fe), and include sub-grid diffusion of these elements in gas via turbulence \citep{Hopkins2016,Su17,Escala18}. Star formation occurs in gas that is locally self-gravitating, sufficiently dense ($ > 1000$ cm$^{-3}$), Jeans unstable and molecular (following \citealt{Krumholz_2011}). Locally, star formation efficiency is set to $100\%$ per free-fall time, i.e., $\rm SFR_{\rm particle} = \textit{m}_{\rm particle} \cdot \textit{f}_{\rm mol} \, / \, \textit{t}_{\rm ff}$ with gas particles stochastically converted to stars at this rate \citep{Katz1996}. Note that this does {\em not} imply that the global efficiency of star formation is $100\%$ within a giant-molecular cloud (or across larger scales). Self-regulated feedback limits star formation to $\sim$1-10\% per free-fall time \citep{FG13,Hopkins17_2,Orr_2018}.

Most of this paper relies on a detailed analysis of five Milky-Way-mass galaxies, which are summarized in the top section of Table \ref{tab:one}.  These zoom simulations are initialised following \citet{Onorbe14}. Three of these galaxies (with names following the convention \texttt{m12$\bullet$}) are isolated and part of the Latte suite ~\citep{Wetzel16,Garrison-Kimmel17,Hopkins17_2,Garrison-Kimmel19}. Two, with names associated with the famous duo (\Romeo~and \Juliet), are part of the ELVIS on FIRE project \citep{Garrison-Kimmel19,Garrison-Kimmel19_2} and are set in Local-Group-like configurations, as in the ELVIS suite \citep{Garrison-Kimmel14}. This suite includes three simulations in total, containing two MW/M31-mass galaxies each. The main haloes were selected so that they have similar relative separations and velocities as of the MW-M31 pair in the Local Group (LG). Table \ref{tab:one} lists the initial baryonic particle masses for each simulation. Latte gas and star particles have initial masses of $7070\, \msun$, whilst ELVIS on FIRE has $\approx 2 \times$ better mass resolution ($m_{\rm i} = 3500\, \msun$). Gas softening lengths are fully adaptive down to $\simeq$0.5$-$1 pc. Star particle softening lengths are $\simeq$4 pc physical and the dark matter force softening is $\simeq$40 pc physical. The last set of seven galaxies is used only in Section \ref{sec:sample_trend} to demonstrate sample-wide trends.

\subsection{Definitions}
\label{sec:define}


\begin{figure*}
    \centering
	\includegraphics[width=0.95 \textwidth, trim = 50.0 0.0 80.0 0.0]{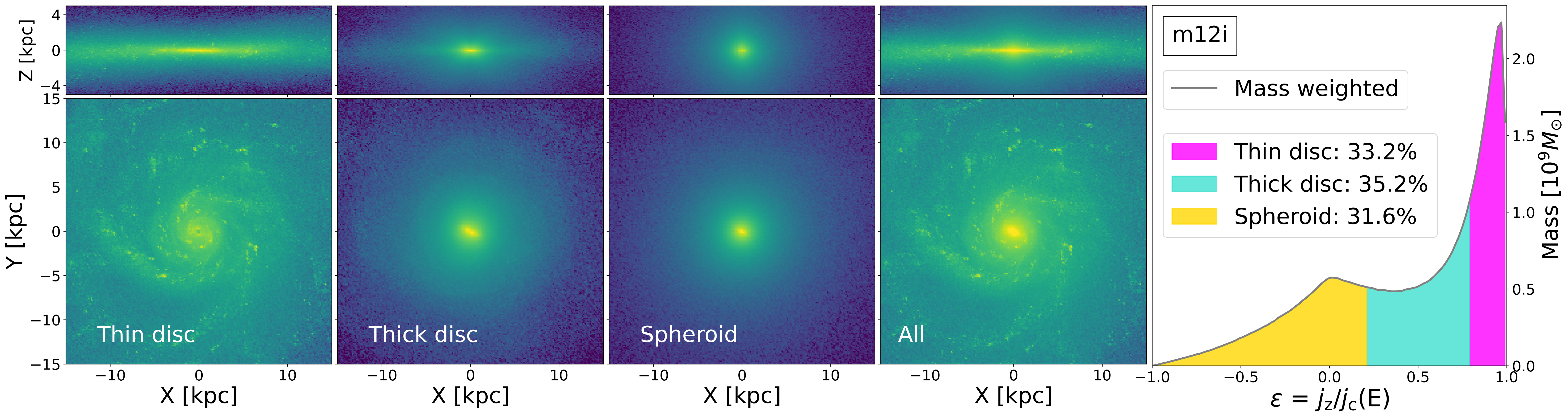}
    \caption[]{Definitions and associated morphologies of different components of \texttt{m12i} at $z = 0$. The left four panels show the edge-on (top) and face-on (bottom) views (2D density weighted by Sloan \textit{r} band luminosity) of the stars we classify as thin disc, thick disc, spheroid, and all stars. The right panel shows the mass-weighted distribution of stellar circularities ($\epsilon$) for all ({\em in situ}) stars within $R_{90}$ ($12.8$ kpc) at $z = 0$. The magenta block marks what we define as thin-disc stars, with $\epsilon\geq 0.8$. The cyan block shows our definition of thick-disc stars, which we set to be those with $0.8 >\epsilon\geq 0.2$. The yellow block marks spheroid stars, which we define as those with $\epsilon< 0.2$. The percentage of stars in each block is shown in the legend. We see that these definitions produce components that qualitatively resemble geometrically-defined discs and spheroids. Note that the inner $\sim 1$ kpc of the ``thick-disc'' component does contain some stars that would likely be identified as a bar or perhaps pseudo-bulge material in a more detailed study.   
    }
	\label{fig:m12i_define_morph}
\end{figure*}

This analysis focuses on {\em in situ} stars that were born within 10\% of the virial radius\footnote{We define virial radius using the \citet{Bryan98} definition. Our results are not sensitive to this specific choice.  Using a fixed value of $10$ kpc for defining the {\em in situ
} stars we study yields similar results.} of the most massive progenitor of each galaxy over time. 
As in Yu et al.\ (\citeyear{Yu2021}, hereafter Y21), we classify star particles  using their orbital circularity, $\epsilon = j_z/j_c(E)$, defined as the ratio of each particle's angular momentum in the $\hat{z}$ direction to that of a circular orbit with the same energy \citep[e.g.,][]{Abadi03}. The angular momentum direction $\hat{z}$ is set by total stellar angular momentum within $10$ kpc of each galaxy's center at the lookback time of interest.~\footnote{In some cases, we use the angular momentum direction at $z=0$. In others, when we track evolution over time, we use the snapshot immediately following the star's birth (not $z=0$). We specify these differences in the text.}  A star with $\epsilon = 1$ is on a circular orbit in the plane of the disc;  $\epsilon < 0$ implies counter-rotation.
We categorise star particles with $\epsilon = 0.8 - 1$ as {\it thin-disc} stars, those with $\epsilon = 0.2 - 0.8$ as {\it thick-disc} stars, and those with $\epsilon = -1.0 - 0.2$ as {\it spheroid} stars. As we show below, we find that circularity appears to be a useful parameter for keeping track of the dynamical evolution of stars over time.

The images shown in Figure \ref{fig:m12i_define_morph} illustrate how these classifications manifest morphologically.  While we have chosen to classify these components using familiar names, it is important to emphasize that those names are traditionally assigned to populations using morphological decomposition rather than dynamical assignment.  Nevertheless, as can be seen in the left-most image, our {\it thin-disc} component does indeed resemble thin discs as usually conceived \citep[see][where we discuss scale heights, etc.]{Yu2021}. Our {\it thick-disc} component consists of a clear vertically-extended disc-like structure, again in keeping with traditional expectations for that name.  However, it also contains a central, bright, and mildly rotating distribution of stars that qualitatively resembles the observed characteristics of pseudo-bulges \citep{Kormendy04}.   Our {\it spheroid} stars (third from left image) have an isotropic configuration that bare qualitative resemblance to a classical bulge and/or inner stellar halo component \citep[e.g.,][and references therein]{Gao20}.  While we make no direct comparisons to observations in what follows, it is worth keeping in mind that our {\it spheroid} component is most aptly associated with an isotropic stellar component that could contain some stars that would be associated with a classical bulge and/or inner stellar halo.

While the orbital circularity $\epsilon$ works decently as a parameter for classification, we also examine two additional parameters in order to understand how different aspects of the kinematics change over time:
the 3D orbital circularity, $\epsilon_{\rm 3D} = j/j_c(E)$, and alignment angle, $\theta = {\rm arccos}(j_z/j)$. 
The 3D circularity $\epsilon_{\rm 3D}$ is defined as the ratio of each particle's total angular momentum to that of a circular orbit with the same energy. Since a circular orbit has maximal angular momentum for a given energy, 3D circularity $\epsilon_{\rm 3D}$ ranges between $0$ and $1$, with $\epsilon_{\rm 3D} = 1$ corresponding to perfectly circular orbits, and $\epsilon_{\rm 3D} = 0$ to purely radial orbits. 
Alignment angle $\theta$ is defined as the angle between each particle's angular momentum to the rotation axis of the galaxy. It describes how aligned the orbit is with respect to the galactic disc. $\theta = 0\degree$ corresponds to orbit that lies perfectly in the disc plane, $\theta = 90\degree$ means that star has a orbit perpendicular to the disc plane, and $\theta = 180\degree$ indicates that it is counter rotating.

We measure the approximate ``birth'' circularity and 3D circularity of each star particle using a post-process analysis of snapshots saved from the simulation.  Specifically, we define $\epsilon_{\mathrm{birth}}$ and $\epsilon_{\rm 3D \, birth}$ at the first snapshot available after each star particle is formed.  The time spacing between snapshots ranges from 16-25 Myr, which is small compared to the timescales of interest ($>100$ Myr).   Note that the angular momentum direction $\hat{z}$ of the galaxy (which affects $\epsilon_{\mathrm{birth}}$ but not $\epsilon_{\rm 3D \, birth}$) is set by the total stellar angular momentum within $0.1R_{\mathrm{vir}}$  at the time of the snapshot immediately following the star's birth (not $z=0$). This method has been verified to produce a steadily evolving reference frame that changes from snapshot-to-snapshot typically on the order of a degree or less in orientation \citep{Gurvich2022}.

We also find it useful to define a transition time between an early bursty phase of star formation and a later steady phase of star formation. We define the bursty phase to end at a lookback time $t_{\rm B}$ when the standard deviation in ``instantaneous'' star-formation rate  first falls below $B=0.2$ times the time-averaged star formation rate:
\begin{equation}
  \frac{\sigma_{10}(t_{\rm B})}{{\rm SFR}_{500}(t_{\rm B})} \equiv B.
  \label{eq:def}
\end{equation}
Here, the ``instantaneous'' star-formation rate is defined to be the rate measured over $10$ Myr intervals and the time-average rate is measured over a $500$ Myr interval, as in \citetalias{Yu2021}.  We use this definition to assign a specific bursty-phase timescale to each galaxy's star formation history. \citet{Gurvich2022} uses the running scatter in the SFH in 300 Myr windows, $\sigma_{\rm 300\ Myr}(\rm log(\rm SFR))$, to quantify the fluctuations and define $t_{\rm B}$ as the time after which this quantity remains below 0.3 dex. The difference between different definitions of burstiness results in variations in $t_{\rm B}$ of a few hundreds of Myr, which is relatively small compared to the cosmological timescales of the transition (see the $t_{\rm B}$ column in Table~\ref{tab:one}).

Finally, we quantify inner CGM virialisation using the ratio of the cooling time of shocked gas $\tcoolsh$ to the free-fall time $\tff$ at an inner radius $r = 0.1~R_{\rm vir}$.  This parameter was introduced by \citet[][]{Stern20}, who used it to show that the bursty to steady transition in galaxy star formation coincides with virialisation of the inner CGM\footnote{See Stern et al. (2021) for a detailed discussion of how we evaluate the ratio $\tcoolsh / \tff$ in our simulations. In short, the cooling time in this ratio is a proxy for the expected cooling time of a hot virialized phase, which is not present before the CGM actually virializes, so the exact definition can be important to reproduce our results.}.
When $\tcoolsh / \tff \gtrsim 1$, the inner CGM is smooth and largely supported by thermal pressure. In contrast, when $\tcoolsh / \tff \lesssim 1$, the inner CGM has large pressure fluctuations and is highly dynamic. The bursty-to-steady transition, as well as a transition from thick-disc to thin-disc formation, coincides with the time when the ratio first crosses $\tcoolsh / \tff \gtrsim 2$ \citep{Stern20,Yu2021}. 

\begin{figure}
	\includegraphics[width=0.99 \columnwidth, trim = 0.0 0.0 0.0 0.0]{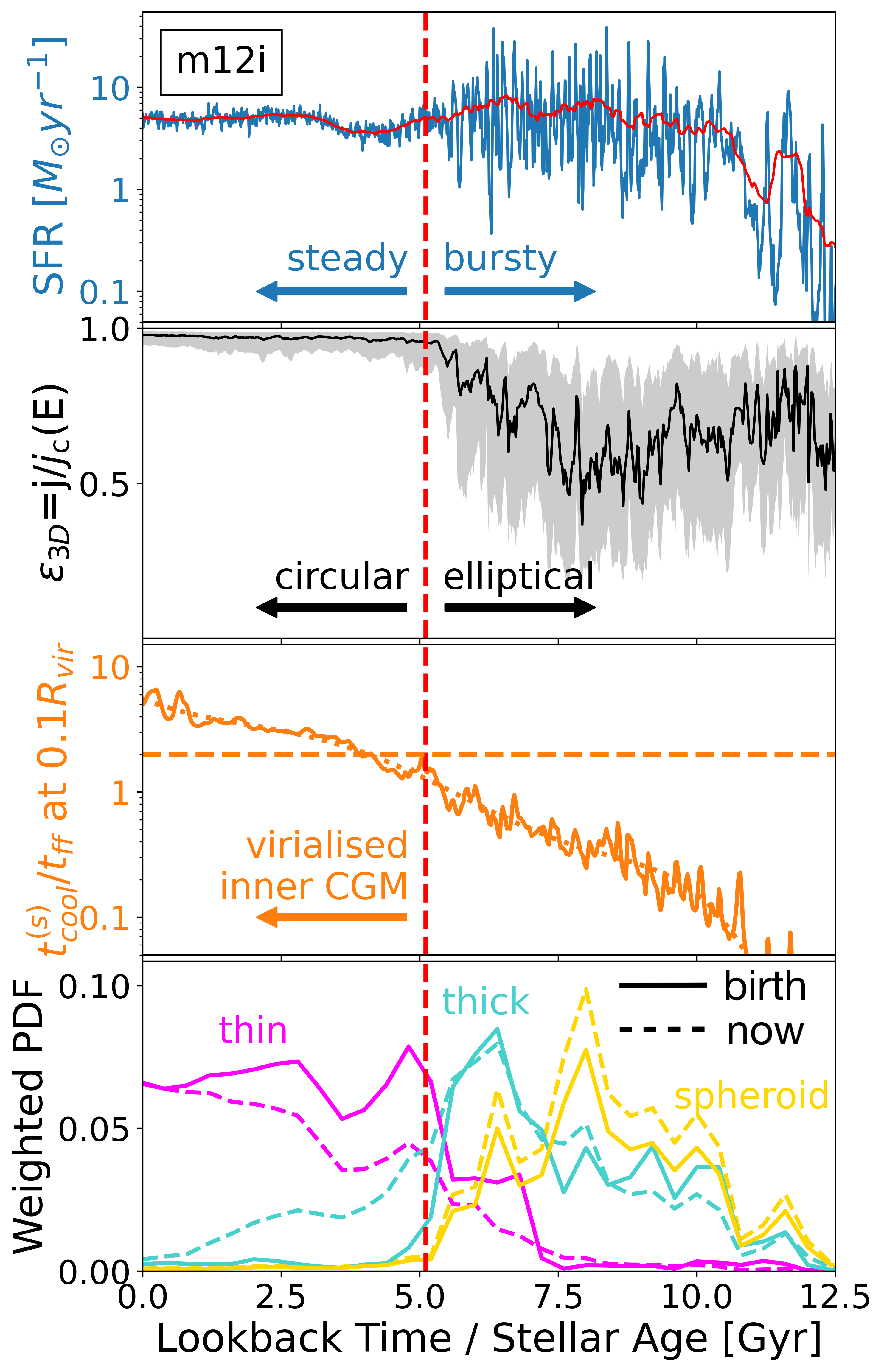}
    \caption[]{Co-evolution of various properties for \texttt{m12i} as a function of lookback time and stellar age. 
    {\bf First Row:} Star formation rate in the galaxy as a function of lookback time. The blue lines show the ``instantaneous'' star formation rate averaged over 10 Myr bins, while the red lines show the ``smoothed'' star formation rate averaged over 500 Myr bins. 
    {\bf Second Row:} The median of 3D orbital circularities, $\epsilon_{\rm 3D} = j/j_c(E)$, of stars younger than 100 Myr as a function of lookback time. The shaded region plots the 16th to 84th percentile range. We see that stars formed during the bursty phase tend to be born on less circular, more ellipital orbits, with a fair amount of variance.  Stars formed during the steady phase are born on extremely circular orbits, narrowly peaked near $\epsilon_{\rm 3D} =1$. 
    {\bf Third Row:} the cooling time to free-fall time ratio measured at $0.1~R_{\mathrm{vir}}$. This parameter was introduced by  \citet[][]{Stern20}, who used it to show that the bursty to steady transition in galaxy star formation coincides with virialisation of the inner CGM. The thin dashed line is a smoothed fit to the thick line measured in the simulations.  When $\tcoolsh / \tff \gtrsim 2$ (horizontal dashed line) the inner CGM is smooth and largely supported by thermal pressure. In contrast, when $\tcoolsh / \tff \lesssim 2$, the inner CGM has large pressure fluctuations and is highly dynamic. 
    {\bf Bottom Row:} Age distribution of stars that have orbital circularities classified thin disc (magenta), thick disc (cyan), and spheroid (yellow). The solid lines show the distribution for stars classified by their circularity at birth, $\epsilon_{\mathrm{birth}} = j_z/j_c(E)$.  The dashed lines show the distribution of stars classified by circularity measured at $z = 0$. The vertical dashed red line indicates the start of the steady phase in star formation. After this time,  stars form on very circular orbits in a thin-disc configuration, and the inner CGM is virialized.  The dashed cyan and magenta lines show that some of the stars that form in the steady phase are heated enough to be classified at thick disc at $z=0$.  However, heating appears to be a secondary effect because the age distributions (dashed) show a similar peak as the formation time distributions (solid). Only the $\sim 10\%$ youngest thick disc stars,  were formed ``thin" in the steady phase.
    }
	\label{fig:m12i_define_param}
\end{figure}

\begin{figure*}
    \centering
	\includegraphics[width = 1. \textwidth, trim = 0.0 0.0 0.0 0.0]{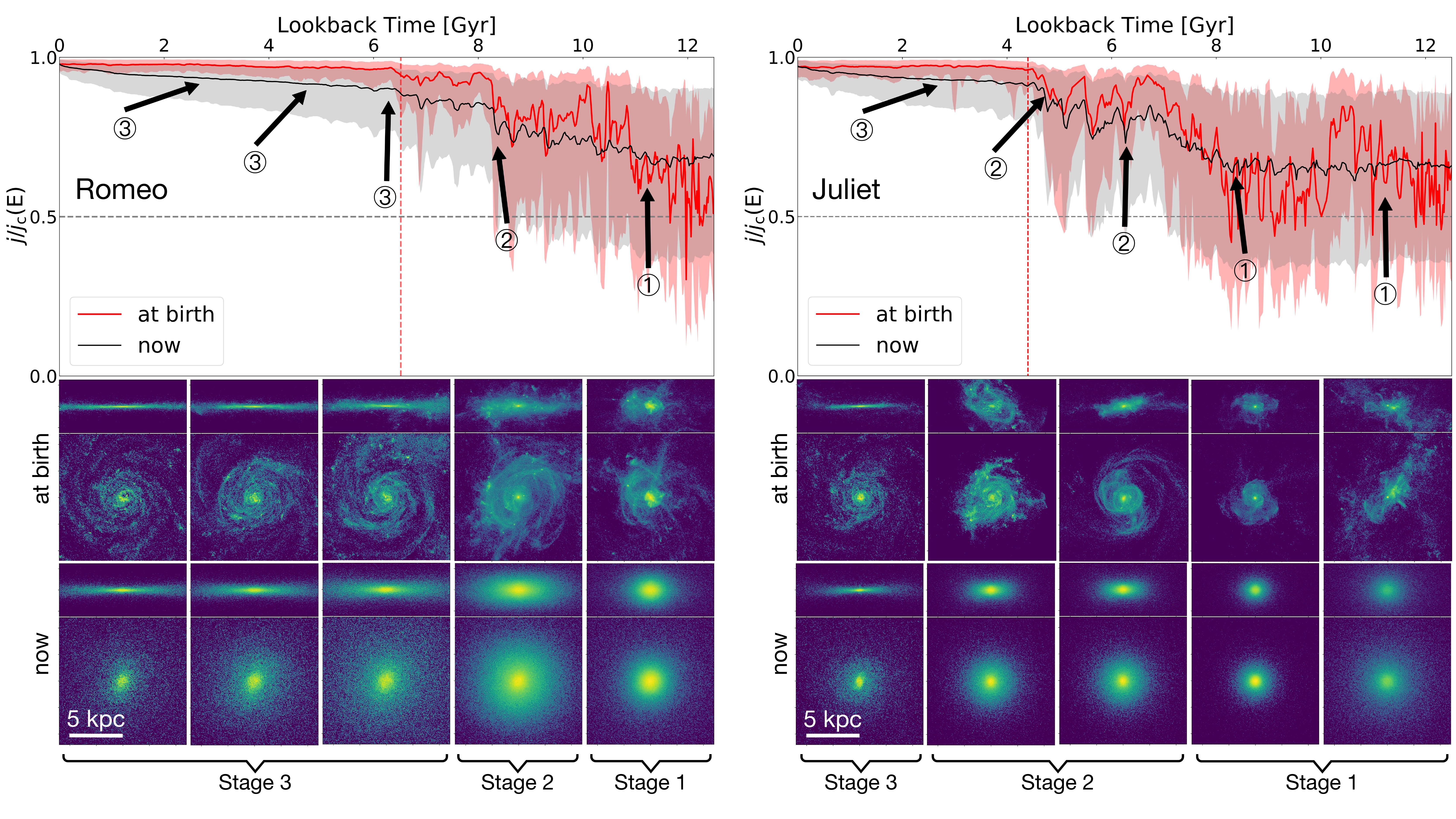}
    \caption[Young-star orbits and morphologies across cosmic time for \texttt{Romeo} and \texttt{Juliet}]
    {
    Young-star orbits and morphologies across cosmic time for \texttt{Romeo} ({\bf left}) and \texttt{Juliet} ({\bf right}). {\bf Top:} The median and $68\%$ range of 3D orbital circularities $\epsilon_{\rm 3D} = j/j_c$ of stars younger than 100 Myr as a function of lookback time. The vertical red dashed line shows the bursty phase lookback time. The solid red line (median) and red-shaded area (one sigma) show the 3D circularities of stars at the time of their birth, while the black line and the grey area show the same quantities for the same stars at $z=0$.  Note that $\epsilon_{\rm 3D}$ is the ratio of the {\em total} specific angular momentum (with no dependence on direction) in units of the circular angular momentum at the same energy: $0$ is purely radial and $1$ is perfectly circular. {\bf Middle:} Luminosity-weighted images, both edge-on (top) and face-on (bottom), for the youngest population (formed within 100 Myr) at five different lookback times -- from left to right: 2.7, 4.7, 6.3, 8.4, and 10.3 Gyr. The arrows in the top panel indicates these times. {\bf Bottom:} Luminosity-weighted images, both edge-on (top) and face-on (bottom), at $z=0$ for the same stars shown in the middle panels. At early times, stars form on more radial orbits and show clumpy/disordered structures. Their spatial distribution today resembles an isotropic spheroid. At late times, stars form on circular orbits and show strong coherence. They remain in relatively thin configurations at $z=0$ as well.
    Note that \texttt{Juliet}'s star formation settled down much later than \texttt{Romeo}'s. Only in the 2.7 Gyr images, after the steady phase has started, do young stars show thin-disc like morphology. The Stage 1, Stage 2, and Stage 3 labels correspond to three phases of evolution we see for all galaxies in our sample: a chaotic bursty phase (1), a spin-up/bursty-disc  phase (2), and a thin-disc phase (3). See text for a more detailed discussion.
    }
	\label{fig:Romeo_Juliet_jjc_morph_evolve}
\end{figure*}

\section{Results}
\label{sec:results}


\subsection{Co-evolution}
\label{sec:co_evolve}

As mentioned in the introduction, FIRE-2 galaxies display co-evolution in a number of parameters related to star formation activity, stellar kinematics, and inner CGM properties. Figure \ref{fig:m12i_define_param} provides an illustrative example of this co-evolution in \texttt{m12i}. Similar figures for the four other primary simulations are provided in Figure \ref{fig:sfh_jjc_CGM_evolve_all}. 

The top panel in Figure \ref{fig:m12i_define_param} shows the star formation history of \texttt{m12i} as a function of lookback time. The star formation rate  (SFR) displayed is averaged over both a short timescale of $10$ Myr (SFR$_{10}$, blue) and a longer timescale of $500$ Myr (SFR$_{500}$, red). 
The relative variance is much larger at early times than at late times.  This is consistent with previous examinations \citep[e.g.,][]{Muratov15,Sparre17,Stern20,Jose2020,Hafen2022,Gurvich2022} that have shown that star formation in massive FIRE galaxies tends to transition from bursty to steady as we approach the present day. The vertical dashed marks the bursty-phase lookback time as defined in the previous section (Eq. \ref{eq:def}), and as described in \citetalias{Yu2021}. We refer to times later than this transition as the ``steady phase'' and times prior to this transition as the ``bursty phase''

The second panel in Figure \ref{fig:m12i_define_param} presents the 3D circularity, $\epsilon_{\rm 3D} = j/j_c(E)$, of newly formed stars (ages < $100$ Myr) as a function of lookback time.  The black line marks the median value while the shaded region shows the $16$th to $84$th percentile range. Note that the circularity distribution of young stars has a sharp transition once the steady phase begins, at a lookback time of $t_{\rm B} \simeq 5$ Gyr.  During the steady phase, stars are born on quite circular orbits, very close to $\epsilon_{3D} = 1$. About 2.5 billion years before the steady phase begins, we see a gradual ``spin up'' phase, where the orbits of young stars become more circular.  Prior to a lookback time of about 7.5 billion years, young stars are born on fairly elliptical orbits.  We see qualitatively similar behaviors for all the galaxies in our sample.  Note that at early times, the 3D circularities of young stars have typical values near $\sim 0.7$.  This is typical of median $j/j_c(E)$ values one often finds for isotropic orbits in spherically symmetric (non-rotating) systems \citep[e.g.][]{vdB99}. 

The third panel in Figure \ref{fig:m12i_define_param} presents the evolution of the ratio $\tcoolsh / \tff $, which tracks the propensity of the inner CGM to be virialized. The ratio  $\tcoolsh / \tff = 2$ is marked.  At early times, $\tcoolsh/\tff \ll 1$, the inner CGM is clumpy and dominated by the supersonic infall of cold gas.  At late times, $\tcoolsh/\tff \gtrsim 2$, and the GCM becomes hot, smooth, and largely supported by thermal pressure. This transition also coincides with the time that the star formation transitions from bursty to steady {\em and} the time when stars begin to form on very circular orbits.  The circularity of young stellar orbits is  enabled by the ability of accreting gas to become coherently aligned in angular momentum space prior to deposition into the galaxy only after the inner CGM becomes smooth and hot \citep{Hafen2022}.

The bottom panel in Figure \ref{fig:m12i_define_param} presents the age distributions of different components identified using circularity $\epsilon = j_z/j_c(E)$. The solid line shows the distribution for the stars classified using birth circularity $\epsilon_{\mathrm{birth}}$.  We find that almost all the thick-disc and spheroid stars form in the earliest periods of galaxy assembly, whilst thin-disc stars form later after the star formation settles down. The difference between thick-disc and spheroid stars is subtle; overall, our identified bulge stars  are a bit older than the thick-disc stars.  The distribution of thick-disc stars peaks near the time of transition, while bulge star ages peak $\sim 2.5$ Gyr prior.

The dashed lines show the distribution of different populations classified using the circularity $\epsilon$ measured at $z=0$, similar to the method adopted in \citetalias{Yu2021}. The two bulge distributions are almost the same. For thick-disc stars, the truncation of the age distribution is more abrupt when using $\epsilon_{\mathrm{birth}}$, while there is an extended tail towards younger age for population classified by $\epsilon$. This is likely due to secular disc heating effects that allow stars born with thin-disc like orbits ($\epsilon_{\mathrm{birth}}>0.8$) into the thick-disc regime (with $\epsilon = 0.2 - 0.8$). The overall time sequence, from bulge formation to thick-disc formation to thin-disc formation, stays the same. While there is some degree of disc heating, as we quantify in Section~\ref{sec:circ_evolve}, the effect is relatively small compared to the birth-orbit trend.

The main takeaway from this subsection is that levels of star-formation burstiness, inner CGM virialization, and birth circularities of new stars appear to be coupled across time.  Figure \ref{fig:sfh_jjc_CGM_evolve_all} shows a similar result for the other four galaxies in  our main sample. At early times, stars are formed in irregular / more radial orbits; the star formation rate is quite bursty, and the inner CGM is clumpy, cool, and not virialized.


\begin{figure}
    \centering
    \includegraphics[width=0.95 \columnwidth, trim = 100.0 0.0 150.0 0.0]{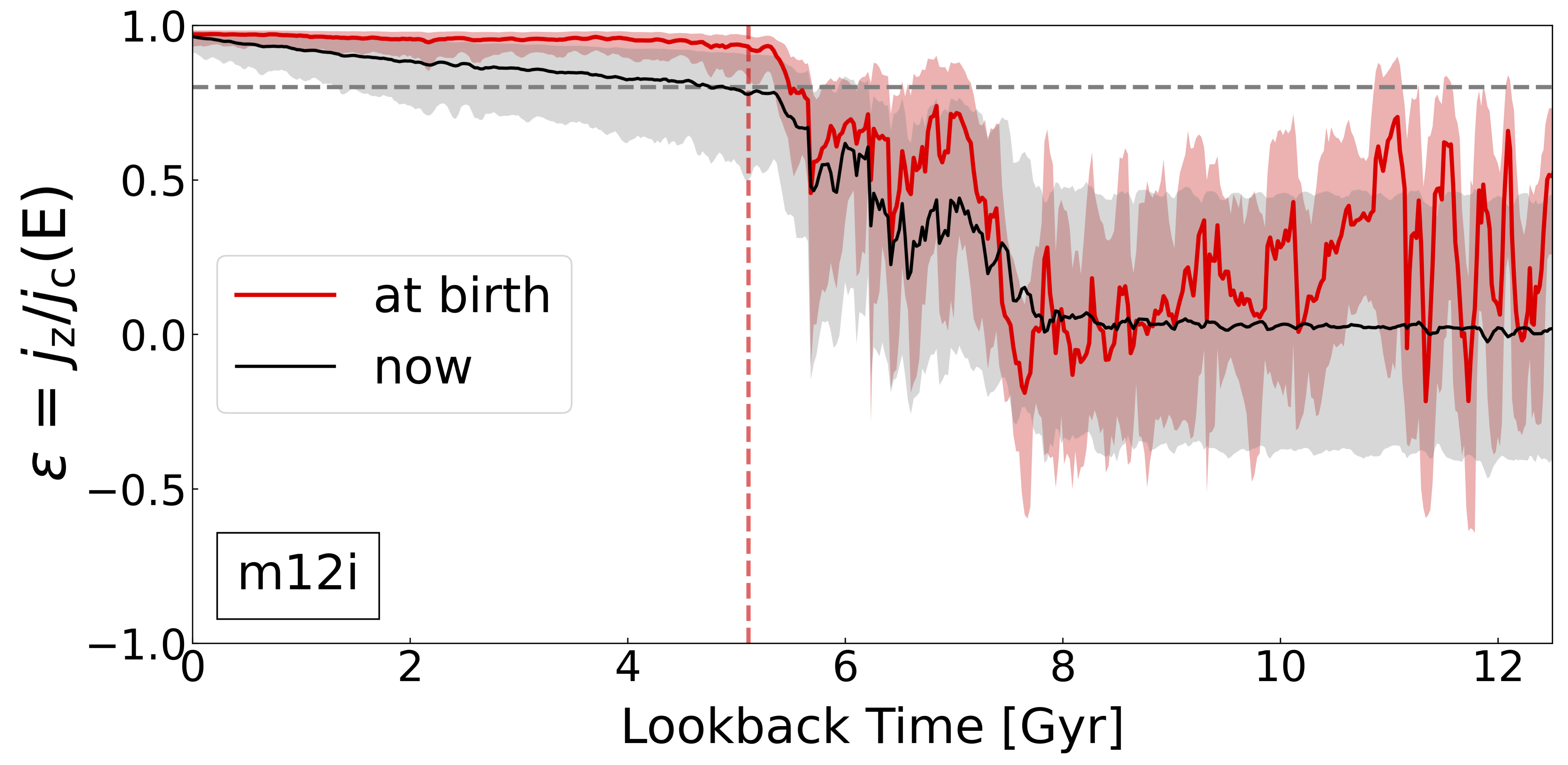}
    \caption[]{
    The median and 68\% (16th-84th percentile) range of orbital circularity of young stars ($< 100$ Myr)  as a function 
    of lookback time for \texttt{m12i}. Unlike the 3D circularity ($j/j_c$) shown in the upper panels of Figure \ref{fig:Romeo_Juliet_jjc_morph_evolve}, the circularity shown here ($j_z/j_c$) encodes information on the shape of stellar orbits {\em and} their orientation with respect to the evolving galaxy's total angular momentum. The vertical red dashed line marks the bursty phase lookback time $t_{\rm B}$. The solid red line and shaded area represent the median and one-sigma distributions of $\rm \epsilon$ at birth, while the the black line and the grey areas are the same quantities for the same stars at $z=0$. For reference, the horizontal grey dashed line marks the threshold above which stars are classified as thin-disc stars. At early times, stars tend to be born with spheroidal-type orbits ($\epsilon < 0.2$) and to evolve into orbits that are similarly radial with respect to the $z=0$ disc ($\epsilon \sim 0$ in the median). At late times, after the bursty phase, stars form on highly aligned, thin-disc like orbits (red).  Those thin-disc stars do get heated over time (grey), but in the median, stars formed after the bursty phase remain in a thin disc. A transition between these two extremes -- a ``spin-up" or thick-disc phase -- occurs between $\sim 5$ and 8 Gyr ago.  During this time, young stars begin to show an increasing level of aligned/coherent rotation, though without the very tight thin-disk alignment we see after the bursty phase has ended.}
	\label{fig:m12i_jzjc_evolve}
\end{figure}

\subsection{Morphology with time}
\label{sec:morph_evolve}

The red regions in the top panels of Figure \ref{fig:Romeo_Juliet_jjc_morph_evolve} show the time evolution of the 3D orbital circularity, $\epsilon_{\rm 3D}$, of young stars (age < $100$ Myr) as a function of lookback time for \texttt{Romeo} (left) and \texttt{Juliet} (right). The solid red lines are median values at fixed formation time and the shaded regions plot the $16$th to $84$th percentile range.  The black solid lines and shaded regions show for the same quantities for the same stars at $z=0$.   The vertical red dotted line marks the bursty phase lookback time $t_{\rm B}$ in each galaxy.  At lookback times smaller than this, the star formation is steady.  The images in the middle panels show edge-on and face-on images\footnote{2D luminosity-weighted in Sloan \textit{r} band.} of the young ($<100$ Myr) stellar populations at five specific times in the past in each galaxy: 2.7, 4.7, 6.3, 8.4 and 10.3 Gyr.  The images in the bottom panels show the distributions of those same stars today.

For \Romeo, we see that 10.3 Gyr ago, the young stars had an irregular morphology (middle panel, far right) with a range of 3D circularities that, in the median, are fairly radial ($\sim 0.6$). Those stars today are arranged in an isotropic, bulge-like configuration (bottom panel, far right). 
Just $\sim 2$ Gyr later, the young stars have begun to show some coherent rotation and take the form of a clumpy/irregular disc. Those stars today are in the form of a smooth, thick disc, with median 3D circularity $\sim 0.8$ and significant scatter.
At 6.3 Gyr, 4.7 Gyr, and 2.7 Gyr (after the steady phase has commenced), \Romeo's young stars are situated in thin discs (middle, left three panels) and and have 3D circularities tightly peaked at $\epsilon_{\rm 3D} \sim 1.0$.
These stars remain in relatively thin configurations at $z=0$, though there has been some heating/elongation of their orbits over time (the gray shaded band is thicker than the red).  We explore this evolution later in Section \ref{sec:circ_evolve}.

The right panels in Figure \ref{fig:Romeo_Juliet_jjc_morph_evolve} show similar behavior for \Juliet, but shifted later in time. This galaxy ended its bursty phase more recently ($t_{\rm B} = 4.40$ Gyr) than \Romeo~ ($t_{\rm B} = 6.52$ Gyr).  This results in the later emergence of thin-disc formation.  While \Romeo~had a pronounced thin-disc component 6.3 Gyr ago, \Juliet~has no thin disc at that time. Only in the most recent image (2.7 Gyr) does \Juliet~start to have young stars forming in a thin-disc like configuration.

The two sample galaxies we show are representative of our larger simulated sample. Specifically we find that there are three stages in the evolution of the FIRE-2 Milky-Way-mass systems over time: 1) a very early, chaotic bursty phase; 2) a later, quasi-stable, bursty-disk ``spin-up" phase; and 3) a stable, thin-disc, ``cool-down"  phase. The three stages are marked in the bottom of Figure \ref{fig:Romeo_Juliet_jjc_morph_evolve}. Stage 1 is associated with very bursty star formation. Young stars resemble irregular, chaotic systems in Stage 1.  Those stars evolve into an isotropic, classical-bulge-like configuration at $z=0$. Later,  the host galaxy enters Stage 2, a ``spin-up'' phase, where a clumpy and disordered disc begins to emerge. Stars born at this stage end up in a thick-disc configuration at $z=0$.
As star formation settles down, we enter Stage 3, when thin-disc formation occurs. During this time, young stars are forming on extremely circular orbits ($\epsilon_{\rm 3D} \simeq 1.0$). Their morphology today also looks much thinner than stars born in Stage 2. Because of the longer steady star formation phase in \Romeo, it also experiences a much shorter stage 1 and stage 2, resulting in an older bulge, an older thick disc, and also a higher thin-disc fraction \citepalias{Yu2021}. \Juliet, instead, stays in Stage 1 and Stage 2 for a longer time. This results in a smaller thin-disc fraction and a younger thick disc.

\subsection{Kinematics with time}
\label{sec:circ_evolve}

Figure \ref{fig:m12i_jzjc_evolve} now tracks the orbital circularities $\epsilon$ of  newly-formed stars (ages < $100$~Myr) as a function of lookback time for \texttt{m12i}. Unlike $\epsilon_{3D}$, the direction-aligned circularity, $\epsilon$, is sensitive to the orientation of orbits with respect to the disc plane. The red lines show the median value of $\epsilon$ when stars form and the black lines show for the same quantity for the same stars at $z=0$. Shaded regions plot the $16$th to $84$th percentile range. The grey horizontal dashed lines marks the threshold we adopt to identify thin-disc stars. At early times, stars were born with low circularities with large fluctuations, which could be due to both the chaotic, bursty nature of star formation and the swift change in the galaxy's orientation \citep[e.g.][]{Dekel20,Santistevan21}. Their current circularity distribution, after a long time of interaction and evolution, becomes centered around $\epsilon \sim 0$, similar to the distribution of an isotropic, spheroidal component. This corresponds to the early chaotic bursty phase we have discussed in Section \ref{sec:morph_evolve} as ``Stage 1''. 
Approximately $2$ Gyr before star formation settles down, the coherence in spin starts to build up and the median value of $\epsilon$ rises to $\sim 0.4$, with some of the stars already surpassing our threshold for thin-disc kinematics. During this ``Stage 2'' phase, the majority stars are born on orbits that we classify as thick-disc. After the star formation becomes very steady, all the young stars are formed in an extremely coherent manner with $\epsilon \sim 1$. The distribution of their current circularity (gray) is wider, especially for stars that formed at earlier times, and becomes more narrow at late times, indicative of a fairly steady heating rate.  This ``thickening'' appears to be a result of a combination of both dynamical heating and vertical torquing, which we explore in more detail below.

\begin{figure*}
    \centering
	\includegraphics[width=0.98 \textwidth, trim = 0.0 0.0 0.0 0.0]{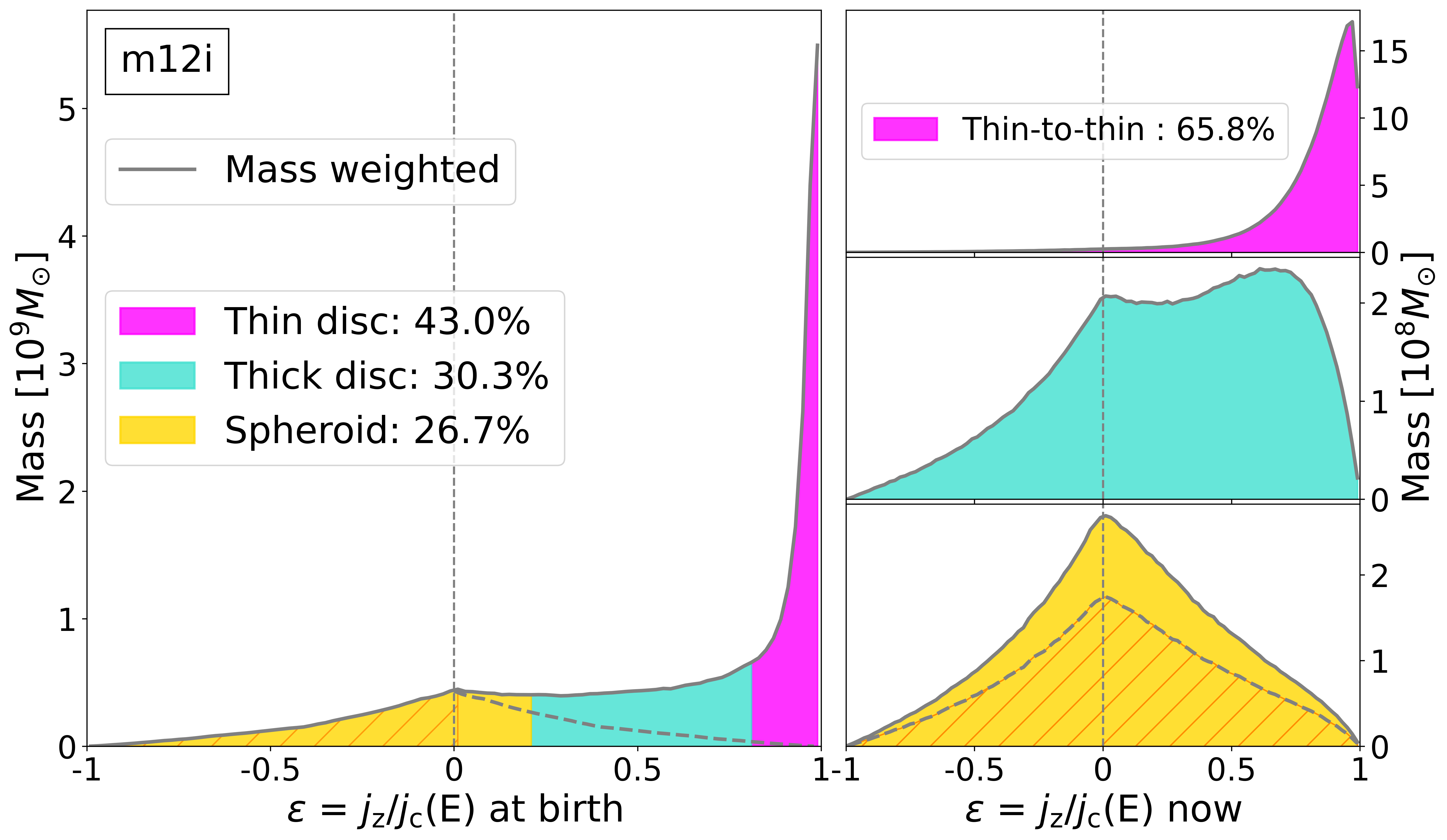}
    \caption[]{How do the orbital classifications of stars evolve from birth to the present day? {\bf Left:} The birth circularity $\epsilon_{\mathrm{birth}}$ distribution for stars that currently reside within ${R_\mathrm{90}}$ ($12.8$~kpc) in \texttt{m12i}. The magenta block marks stars that formed with thin-disc kinematics,  with $\epsilon_{\mathrm{birth}}\geq 0.8$. The cyan block marks thick-disc stars, which we define to be those with $0.8 >\epsilon_{\mathrm{birth}}\geq 0.2$. The yellow block marks spheroid stars with $\epsilon_{\mathrm{birth}}< 0.2$. The mass-weighted fraction of stars in each block is shown in the legend. {\bf Right:} The present-day circularity  distributions for each of the three populations defined by their birth circularities in the left panel.  The colours in each of the right panel match the components they track as defined on the left: thin-disc, thick-disc, and spheroid, from top to bottom.  Note that the vertical (mass-density) axis for each panel is different. {\bf Top Right:} The current circularity distribution for stars born with thin-disc like orbits (magenta). Most of these stars retain thin-disc orbits.  The distribution does have a tail towards lower circularity, which is due to disc heating, but $65.8\%$ of the stars still have current circularity greater than $0.8$. {\bf Middle Right:} The distribution of the stars identified as thick disc (cyan block) in the left panel evolve to have a distribution that looks like a combination of a thick disc and isotropic bulge population.  {\bf Bottom Right:} Stars born with spheroid-like orbits (yellow) evolve into a nearly isotropic distribution at the present day, though there is a slight preference for prograde orbits.  The red-shaded region shows what happens to an isotropic spheroid population that is forced to be symmetric around $\epsilon_{\mathrm{birth}}=0$ (dashed grey, lower left). We see that this choice has a minimal impact on the shape of the distribution today.
    }
	\label{fig:m12i_evolve}
\end{figure*}

Figure \ref{fig:m12i_evolve} illustrates how stellar $\epsilon$ values evolve in \texttt{m12i} in a slightly different way, now with emphasis on our classification categories.
The left panel is similar to the most right panel of Figure \ref{fig:m12i_define_morph}, but now we show the distribution of birth circularity $\epsilon_{\rm birth}$ for all stars in the galaxy, instead of the circularity today. Using our standard definition, stars are classified as having thin-disc orbits at birth (magenta), thick-disc orbits at birth (cyan), and spheroid-like orbits at birth (yellow). The mass-weighted fractions of the different components are listed on the plot.  Note 43 \% of stars are born on thin disc orbits.  This is higher than the fraction of stars that are on thin-disc orbits today (33\%, see Figure \ref{fig:m12i_define_morph}).  This is consistent with the observation that some stars that are born in thin disc orbits become heated over time.

The right panels of Figure \ref{fig:m12i_evolve} display the distribution of current circularities $\epsilon$ for each of the birth-based categories. Stars that are born with thin-disc circularities ($\epsilon_{\rm birth} > 0.8$, magenta) end up in clearly prograde, disc-like configuration today, with an $\epsilon$ distribution that peaks at $0.9$ and with $66\%$ of the distribution still in a thin-disc (with $\epsilon > 0.8$).  There is a tail less well aligned orbits, likely as a result of heating or torquing over time (see below).  For stars born with thick-disc orbits (cyan), their current circularity distribution resembles a combination of a prograde/thick-disc component (peaking at $\epsilon \sim 0.6$) and a broad isotropic/bulge component centered on  $\epsilon \sim 0$.  Stars born with spheroidal orbits (yellow) tend to stay in an isotropic distribution today, with a mild prograde asymmetry.  

Ideally, an isotropic spheroid population would have symmetric circularity distribution around 0.  For the simplicity of this work, we have adopted a sharp cut in circularity to define spheroids ($
\epsilon < 0.2$). We have explored the impact of imposing symmetric distribution, mirroring the observed $\epsilon <0$ distribution (hashed yellow in the left panel) for $\epsilon >0$  (marked by grey dashed line in the left panel).  In this dashed-line selection, we have simply chosen particles randomly at fixed $\epsilon$ such that they inhabit the symmetric distribution shown.  The difference between including the stars with $\epsilon = 0.0 - 0.2$ at birth is minimal since statistically only a small fraction of stars fall in this region (yellow block above the grey dashed line).
Overall, we observed a trend that stars born with bulge-like orbits retain spheroid-like orbits today (dashed grey distribution on the right).

\begin{figure*}
    \centering
	\includegraphics[width=0.98 \textwidth, trim = 0.0 0.0 0.0 0.0]{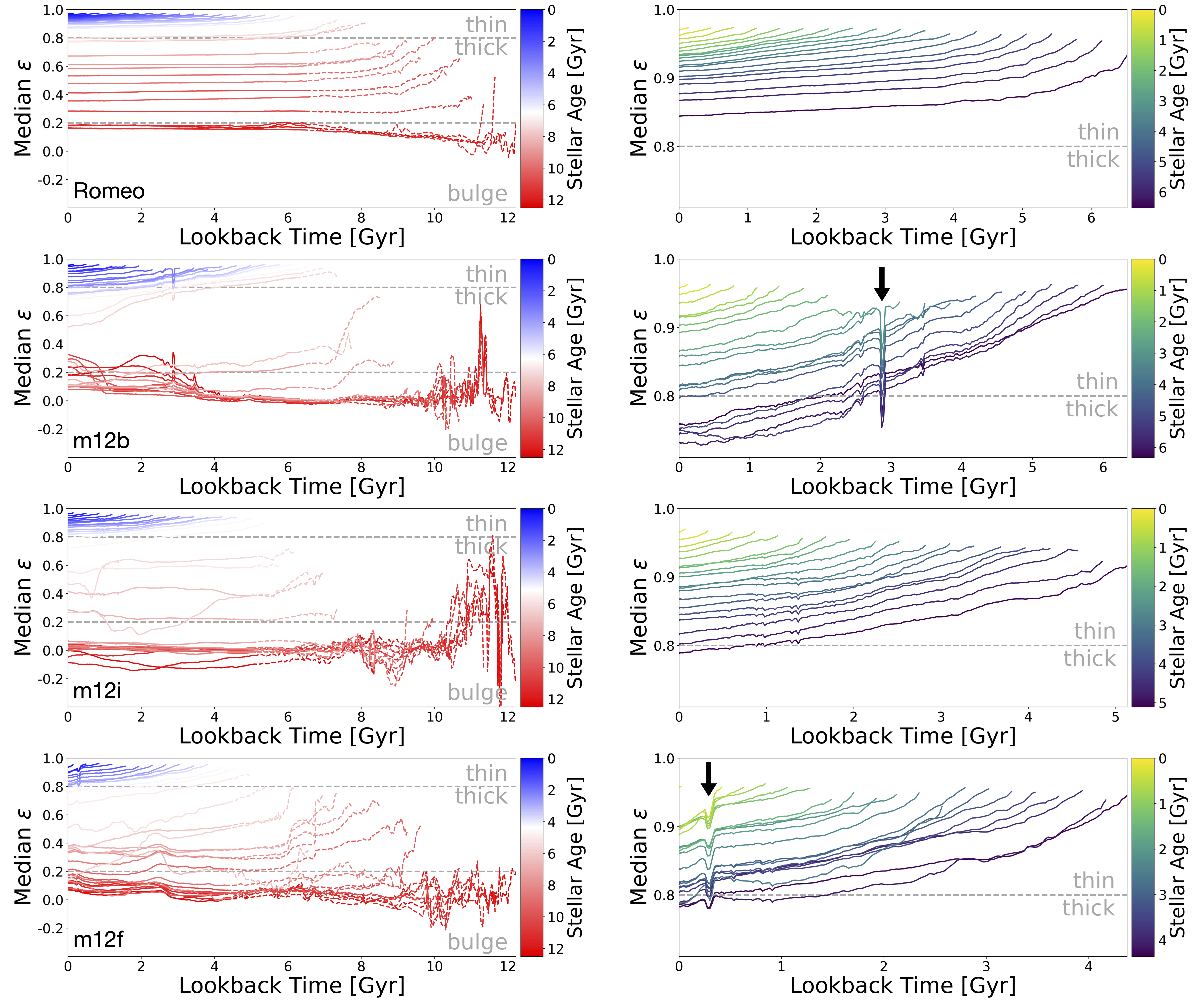}
    \caption[]{
     Evolution of the population-median circularity $\epsilon = j_z/j_c(E)$ of mono-age stellar populations for  \texttt{Romeo}, \texttt{m12b}, \texttt{m12i}, and \texttt{m12f}. Each curve is coloured by stellar age of the population. {\bf Left:} These figures span all of cosmic time. The colour bars are slightly different for each galaxy, such that the bursty-phase transition time is always white.  Red lines track the evolution of mono-age populations that formed during the bursty phase.  Blue lines track the evolution of mono-age populations that formed during the steady phase. The older populations (red lines) tend to evolve to spheroid-like orbits today ($\epsilon \lesssim 0.2$) while younger populations (blue lines) form with much higher circularities ($\epsilon \sim 1$). 
    {\bf Right:} Zoomed-in plots for only the populations that have formed during the steady phase for each host galaxy. Note that the time-axis is slightly different for each galaxy because each galaxy has a different bursty-phase lookback time. The stellar ages are again mapped to the colour bar. The key takeaway is that, in the steady phase, new stars are born with thin-disc orbits and  remain above the 0.8 thin-disc threshold in most cases.  That is, most stars that form during the steady-star-formation period have thin-disc kinematics today. Note that, for \texttt{Romeo} and \texttt{m12i}, all populations form during the steady phase experience an initial drop in $j_z/j_c$ but quickly plateau to nearly nearly constant values. 
    Both \texttt{m12b} and \texttt{m12f} undergo $\sim 1/10$ merger events during the steady phase and this appears to drive slightly different behaviour, with less clear plateauing.  The time of each merger is indicated by an arrow, and there is an obvious dip at the time of merger.  See the text for a discussion of these dips.
    }
	\label{fig:tracking_circ_full}
\end{figure*}

\subsection{Evolution of mono-age populations}
\label{sub:evolve}

Figure \ref{fig:tracking_circ_full} shows the evolution of circularity $\epsilon$ for different mono-age stellar populations in \texttt{Romeo}, \texttt{m12b}, \texttt{m12i}, and \texttt{m12f}, from top to bottom, respectively. Here we have binned star particles based on their birth ages in $100$ Myr increments and have calculated the median value of circularity $\epsilon$ in each age bin as a function of lookback time. Lines are coloured by the stellar age of the population (see colour bars) and are dashed when the host galaxy is in its bursty phase.  They are solid during the steady phase for each galaxy. 

The left column of Figure \ref{fig:tracking_circ_full} tracks the evolution of each mono-age population over all time.  The colour bars are scaled differently for each galaxy in order to emphasize the regularity when normalized with respect to the bursty phase lookback time.  Specifically, colour bars are set so that the white colour corresponds to the transition time from bursty to steady star formation.  Populations that formed during the bursty phase are coloured red and  populations formed during the steady phase are coloured blue. The two grey dashed lines mark our thresholds for the classification of three components (thin disc, thick disc, and spheroid). We observe the same behavior in all four galaxies: stars that form during the early bursty phase (red) mostly have orbits characteristic of spheroid stars ($\epsilon < 0.2$) while populations born during steady phase (blue lines) have median circularities characteristic of thin-disc stars. While the red (early-forming) populations show large fluctuations,  they all approach $\epsilon \sim 0$ (characteristic of an isotropic population) at late times.  Interestingly, there is a brief period in \texttt{m12i} at a lookback time of $\sim 11$ Gyr where the young stars have more coherent, thick-disc like orbits $\epsilon \sim 0.4$; though as in all other cases they evolve to have isotropic orbits with respect to the final system at $z=0$.

The right panels of Figure \ref{fig:tracking_circ_full} are ``zoomed-in'' versions of the left panels and concentrate exclusively on the steady phase. The lines are again coloured by the stellar age. Stars formed during this period are born along a tight plane with extremely circular orbits ($\epsilon\sim 1$).  Their circularities decrease slowly with time, but in almost all cases retain orbits characteristic of thin discs to the present day, with $\epsilon\gtrsim 0.8$ at $z=0$.  The steady-phase mono-age populations for \texttt{Romeo}, \texttt{m12i}, and \texttt{m12f} all show more rapid evolution in orbital circularity within the first $\sim 500-1000$ Myr after being born, and begin to flatten somewhat at late times. This behavior is most extreme for \texttt{Romeo} where the evolution plateaus to near constant circularity at recent lookback times.

The tracks in \texttt{m12b} show somewhat different behavior, with all populations evolving slowly but continuously to lower circularity until the present day.  This is likely because the disc experiences a merger with an LMC-size satellite at a lookback time of approximately 2.9 Gyr (marked by the arrow). This merger, which comes in on a polar orbit,  produces a sharp feature in all of the mono-age tracks (and also drives a small starburst; see \citetalias{Yu2021}). This feature is almost certainly an an artifact of our definition of spin axis. When the merger happens, the satellite galaxy gets close to the host galaxy. This affects the $\hat{z}$ direction  of the galaxy, since we use all the stars within $0.1R_{\mathrm{vir}}$ in calculating the direction of total angular momentum.  While the mass of stars in the accreted object is small, they are all moving coherently at the time of the merger and this changes the bulk angular momentum direction dramatically.  After the merger, the accreted stars become phase-mixed and no longer contribute substantially to the total angular momentum. 

Even with the fairly significant merger in \texttt{m12b}, the median circularity of the oldest post-bursty-phase population remains within our thin-disc classification ($\epsilon > 0.8$). Note that \texttt{m12f} also experiences a similarly-sized merger, at a lookback time of $\sim 300$ Myr (see arrow).  The effect on $\epsilon$ evolution is less dramatic, but still visible: the ``plateauing'' in circularity stops after the merger, and we see a mild drop afterwards. The weaker effect seen with this merger is possibly because the orbit is prograde (though interestingly, this merger drives a larger starburst, see \citetalias{Yu2021}).

\begin{figure*}
    \centering
	\includegraphics[width=0.98 \textwidth, trim = 0.0 0.0 0.0 0.0]{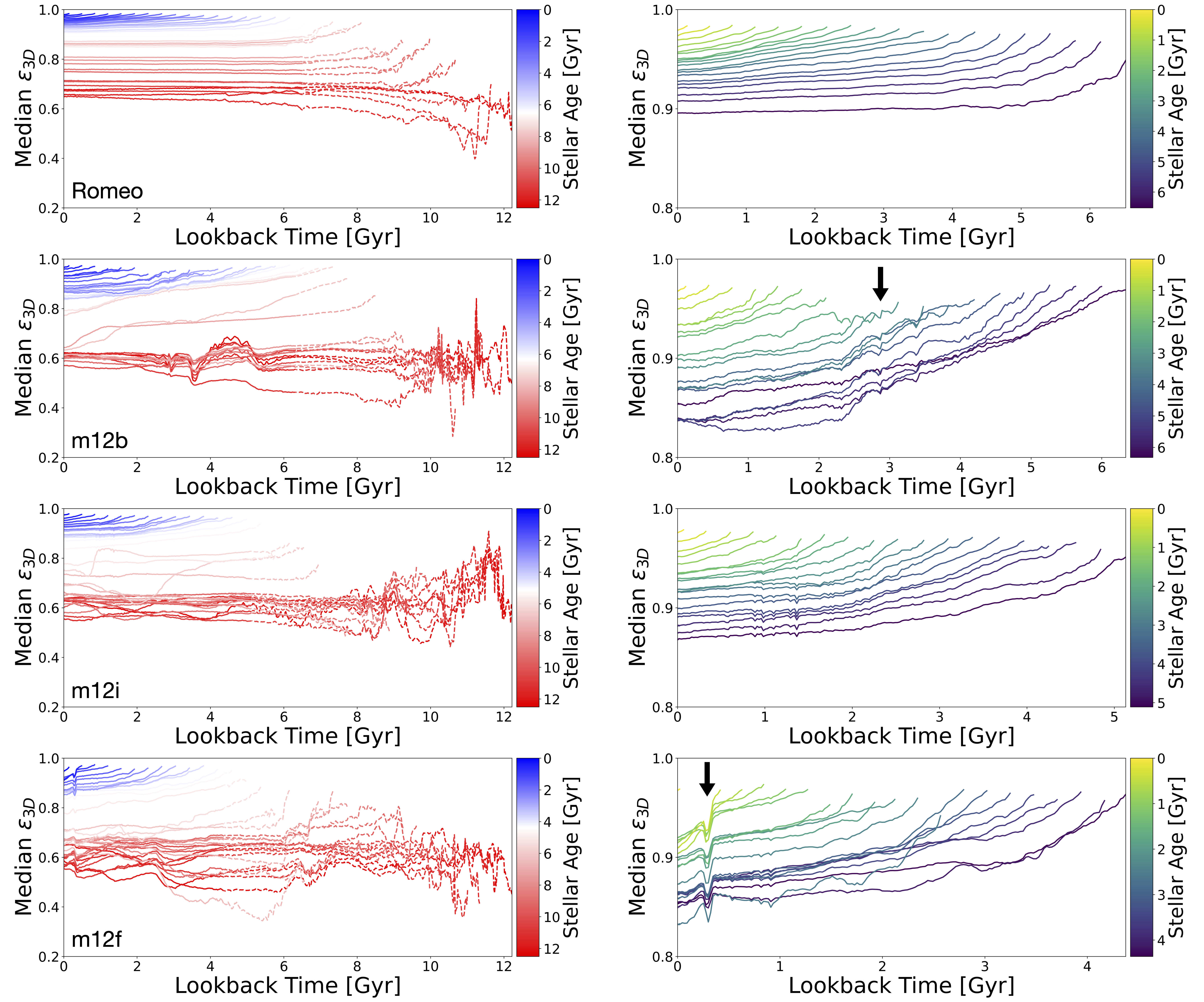}
    \caption[]{Evolution of the median 3D circularity $\epsilon_{\rm 3D} = j/j_c(\rm E)$ for mono-age stellar populations in \texttt{Romeo}, \texttt{m12b}, \texttt{m12i}, and \texttt{m12f}. As in Figure \ref{fig:tracking_circ_full}, each curve is coloured by stellar age of the population. {\bf Left:} Each line tracks the evolution of $\epsilon_{\rm 3D}$ over cosmic time, from birth (right) to the present day (far left).  For each galaxy, red lines correspond to populations that formed during the bursty star-formation phase, while blue lines track stars that formed during the steady phase. The older populations (red lines) tend to form with less circular orbits ($\epsilon_{\rm 3D} \simeq 0.6-0.8$) while younger populations (blue lines) form on quite circular orbits ($\epsilon_{\rm 3D} \sim 1$). Bursty-phase stars tend to remain on fairly radial orbits today.  Steady-phase stars evolve to slightly less circular orbits today but remain more circular than an of the bursty-phase populations.
    {\bf Right:} Zoomed-in plots for only the populations that have formed during the steady phase for each host galaxy. These stars are born on very circular orbits, and become slightly less circular over time, with the sharpest evolution seen within the first $\sim 500$ Myr after formation.  The arrows in \texttt{m12b} and \texttt{m12f} indicate times that these galaxies experience $\sim 1/10$ mergers.  Note that the merger in \texttt{m12b}, which is polar, does {\em not} cause a sharp feature in the 3D circularity of orbits; this suggests that  the distinct feature we see in Figure \ref{fig:tracking_circ_full} is driven by an artifact of how we define the orientation of the galaxy (see text).  Conversely, the merger in \texttt{m12f}, which is prograde, {\em does} cause a sharp feature in 3D circularity, suggesting that this merger has heated existing stars to some extent and driven their orbits to become less circular.  
    }
	\label{fig:tracking_circ3D_full}
\end{figure*}

\begin{figure*}
    \centering
	\includegraphics[width=0.98 \textwidth, trim = 0.0 0.0 0.0 0.0]{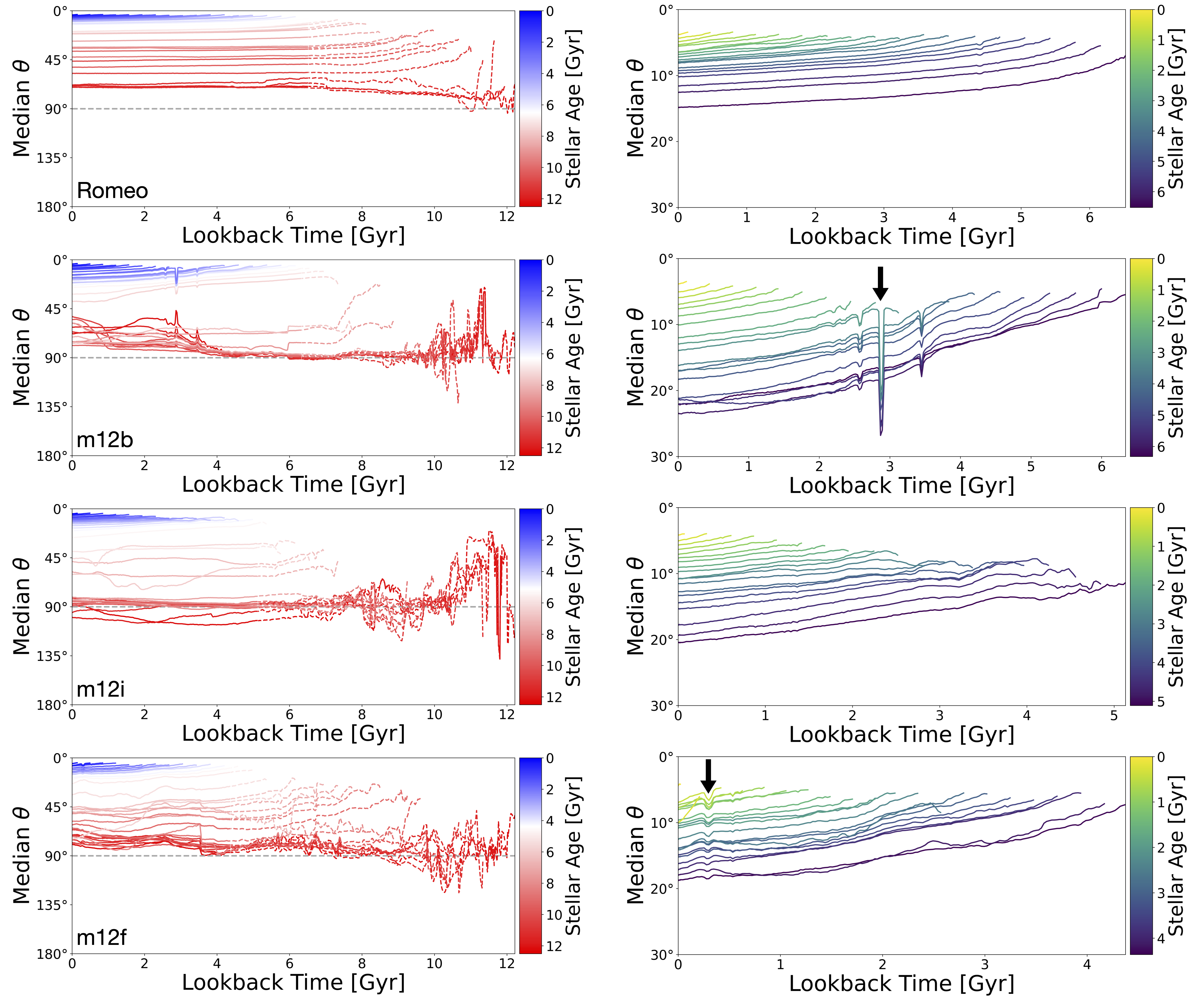}
    \caption[]{Evolution of the median alignment angle $\theta = \arccos{(j_z/j)}$ of mono-age stellar populations with respect to the evolving galaxy's total angular momentum direction in \texttt{Romeo}, \texttt{m12b}, \texttt{m12i}, and \texttt{m12f}. As in Figure \ref{fig:tracking_circ_full}, each curve is coloured by stellar age of the population. {\bf Left:}   The colour bars are set so that all the populations form during the bursty phase are coloured red while the populations form during the steady phase are coloured blue. Orbits of older populations (red lines) tend to be in random directions with respect to the orientation of the galaxy while younger populations (blue lines) form more aligned with the disc plane. 
    {\bf Right:} Zoomed-in plots for only the populations form during the steady phase for each host galaxy. All populations form fairly aligned with the disc ($\theta \lesssim 10\degree$). Soon after formation, most populations experience some evolution initial change in $\theta$, but later plateau at a nearly constant value.  The arrows in \texttt{m12b} and \texttt{m12f} indicate times that these galaxies experience $\sim 1/10$ mergers.  Note that the merger in \texttt{m12b}, which is polar, causes a sharp feature in the relative alignment of stellar orbits; this suggests that  the distinct feature we see in Figure \ref{fig:tracking_circ_full} is driven by an artifact of how we define the orientation of the galaxy (see text).  Conversely, the merger in \texttt{m12f}, which is prograde, does {\em not} cause a sharp feature in orientation of the orbits, suggesting that this merger has driven orbits to be come more elongated (see Figure \ref{fig:tracking_circ3D_full}) but has not affected their orientations significantly.  
    }
	\label{fig:tracking_theta_full}
\end{figure*}

Figures \ref{fig:tracking_circ3D_full} and Figure \ref{fig:tracking_theta_full} complement Figure \ref{fig:tracking_circ_full} by tracking the evolution of 3D circularity $\epsilon_{\rm 3D}$ and the alignment angle $\theta$ of the orbits with respect the existing disc. The 3D circularity $\epsilon_{\rm 3D}$ quantifies how circular the orbits are regardless of orientation with respect to the disc plane.  The alignment angle $\theta$ quantifies the level of alignment of angular momentum in the stars with respect to the rotation axis of the galaxy disc. While $\epsilon$ provides a measure of thickness because it depends on orbital orientation and elongation, these figures provide insight on how the change in circularity $\epsilon$ is driven separately by orbital elongation (regardless of orientation of the orbit) or vertical torquing of the orbit out of the disc plane. 

In the left panels of Figure \ref{fig:tracking_circ3D_full}, we see that stellar populations born during the bursty phase typically start off with fairly radial orbits ($\epsilon_{\rm 3D} \simeq 0.6$). The left panels of Figure  \ref{fig:tracking_theta_full} show that these early-forming stars also have significant misalignment, with median $\theta \sim 90^\degree$ in most cases. Note that this does not mean that most stars are actually orbiting on orbits perpendicular to the disc.  Rather, the distribution of orbital planes is quite random, such that in the median the angle is $\theta \sim 90^\degree$.

As mentioned above, \texttt{m12i} deviates some from the trends we see in the other cases.  In particualr, there is a brief phase at $\sim 11$ Gyr lookback time where populations of young stars have 3D circularities as large as $\sim 0.8$.   At the same time, the median orbital alignments are slightly tighter, with values sometimes as small as $\sim 45^\degree$.  However, the angular momentum direction of the system varies quickly enough that we also see median orbits counter-rotating with $\theta \sim 135^\degree$ over short time intervals.  This level of irregularity in total angular momentum direction at early times helps explain why most of the $\sim 11$ Gyr-old stars in \texttt{m12i} have evolved to inhabit an isotropic ``spheroid'' population at late times.  This can be seen by comparing the red (at birth) and gray (now) distributions over the same time period in Figure \ref{fig:m12i_jzjc_evolve}.

The right-hand panels of Figure \ref{fig:tracking_circ3D_full} and Figure \ref{fig:tracking_theta_full} focus on the evolution of mono-age populations since the bursty phase has ended in each galaxy.  There are a few interesting trends visible here.  First, in all galaxies except \texttt{m12b}, we see that stellar populations stars born with gradually increasing degree of circularity and alignment within the plane as we approach the present day.  That is, even the youngest stars become increasingly thin-disc like as time progresses.  This is not the case in \texttt{m12b}, where the merger (marked by the arrow) resets the trend: post-merger, the orbits steadily become more aligned with the plane and have more circular orbits. 

After birth, the stars' orbits become gradually less circular and more misaligned, though the level heating/torquing is mild and in most cases the mono-age lines plateau to near-constant values after a short $\sim 500$ Myr period just after the stars form, where the elongation/torquing behavior is quickest. Interestingly, a late-time prograde merger in \texttt{m12f} (see arrow) causes a fairly significant drop in 3D circularity but not in the alignment angle. 

The preceding discussion has shown that while there is some secular or merger-driven heating of orbits after formation, the degree of heating is small compared the broader trends with birth orbits over cosmic time.  The scale of the vertical axes in the right column of Figures \ref{fig:tracking_circ_full} -- \ref{fig:tracking_theta_full} is quite narrow compared to those of the left, and in most cases even the first populations to form after the bursty phase ends do not evolve enough to become thick-disc stars by our classification.

\begin{figure*}
    \centering
    \includegraphics[width=0.98 \textwidth, trim = 0.0 0.0 0.0 0.0]{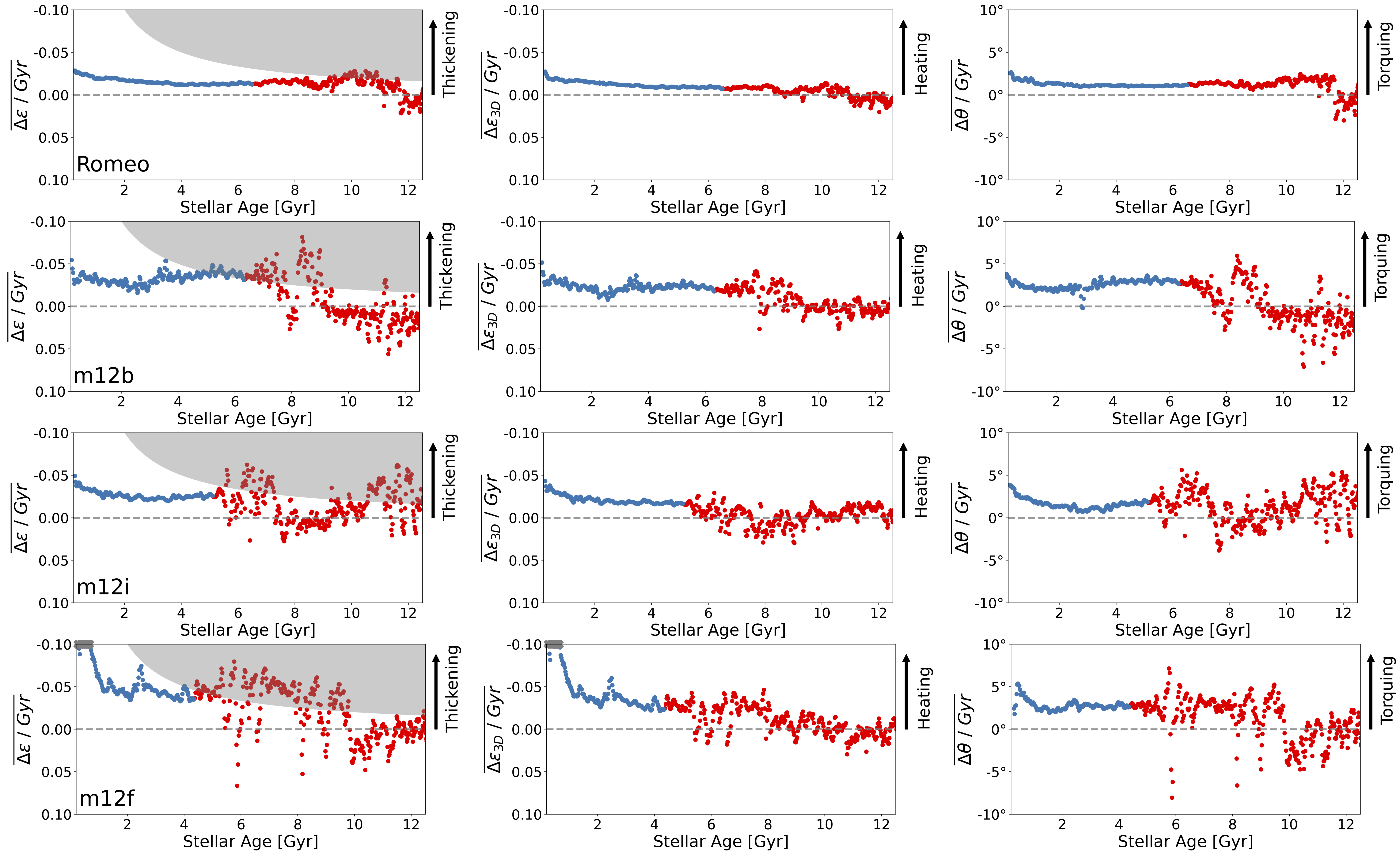}
    \caption[]{ Average change in circularity $\epsilon$ per Gyr (left), 3D circularity $\epsilon_{\rm 3D}$ per Gyr (middle), and alignment angle $\theta$ per Gyr (right) of all mono-age stellar populations for  \texttt{Romeo}, \texttt{m12b}, \texttt{m12i}, and \texttt{m12f}.  Each dot represents the average rate of change for a mono-age population from birth to the present day, computed by dividing the total change in the quantity since birth by the time since birth.  Red dots are populations born during the bursty phase of each galaxy (age>$t_{\rm b}$).  Blue dots show rates for populations born after the star formation settles down. 
    The grey bands in the left panel estimate the amount of ``thickening'' needed for stars born thin ($\epsilon_{\rm birth}=1.0$) to become thick ($\epsilon_{\rm now}\le0.8$). Note that, this region is only relevant as a comparison for population born in steady phase since stars born in bursty phase have $\epsilon_{\rm now} < 0.8$ at birth. The middle panels and right panels show the kinematic heating and vertical torquing rates, respectively. The changes in both quantities are relatively small compared to the trends we observe in birth orbital properties with cosmic time. Note that the blue points rise sharply at the youngest stellar ages in \texttt{m12f}.  This is because this galaxy experiences a late-time merger. The other galaxies show a much milder rise in heating rate for the youngest galaxies.  This is an artifact of our measuring the rate of heating since birth.  All populations show their most dramatic heating within 1 Gyr of formation.  The youngest stars are still in their rapid heating phase when the time-average heating rate is measured. 
    }
	\label{fig:delta_perGyr}
\end{figure*}

Figure \ref{fig:delta_perGyr} explores the question of heating rate in more detail by plotting the average rate of change in orbital properties of all mono-age stellar populations between the time of their birth and today. The left panels show the average change in circularity $\epsilon$ per Gyr, calculated by dividing $\epsilon_{\rm now} - \epsilon_{\rm birth}$ by the stellar age. Red dots represent populations that form during bursty phase and blue dots show the ones born in the steady phase.

The gray shaded region provides a sense of the rate ``thickening'' that would be required to change a star born in a perfectly thin disc ($\epsilon = 1.0$) to one that has just enough orbital circularity to inhabit the thick disc at $z=0$ by our definition ($\epsilon = 0.8$). The gray band rises towards the current epoch because stars born later have less time to be heated and thus need a higher heating rate to join the thick disc before $z=0$. Note that this comparison only makes sense for stars born in steady phase (blue) since stars born during the bursty phase are typically born with thick-disc or spheroid-like orbits already. We see that in almost all cases, the blue points sit below the gray region.  This means that the average ``thickening'' since birth is simply not enough to turn thin-disc stars into thick-disc stars in most cases.

The middle and right columns explore the rate of change in 3D circularity $\epsilon_{\rm 3D}$ and alignment angle $\theta$.  We show these results in order to disentangle the effects of orbital elongation and torquing. As we can see, both mechanisms work to thicken the newly-born stars, but the amount of heating and torquing is minimal. The change in $\epsilon_{\rm 3D}$ is smaller than $0.05$ per Gyr and the change in $\theta$ is smaller than 5\degree~per Gyr.  

By comparing the magnitude of the left and middle columns, we see the thickening rate (left column) very similar in magnitude to the heating rate (middle column), suggesting that heating (rather than torquing) is dominating the thickening evolution.

Note that the average rate of change typically shows an upward trend in stars with the youngest age ($< 1$ Gyr), which is 
driven by the fact that the denominator in the rate calculation is always the time since birth.  We find that the heating rates for stellar populations are usually highest within the first $\sim 500$ Myr of formation. This makes sense in our simulations because star formation occurs only in overdense, self-gravitating structures, where also the strongest dynamical perturbations after formation will occur.
This can be seen in the right columns of Figure \ref{fig:tracking_circ3D_full}, for example.  We further explore these effects in Appendix~\ref{app:2} (Figure \ref{fig:change_in_delta1} and Figure \ref{fig:change_in_delta2}). 

The evolutionary trends we have presented in this section show that stars born on circular, disc-like orbits remain mostly in thin discs today and that stars born on elliptical orbits remain on similar orbits today. Broadly speaking, stars that currently exist in kinematic/morphological classes today were born that way. Though we do see some dynamical heating and torquing over time, the level of these evolutionary effects are secondary compared to the birth properties.  Of course, given our small sample size and the fact that we have focused on Milky-Way size galaxies, our results may not be universal for all galaxy classes. In \citetalias{Yu2021}, we explored the importance of mergers in shaping thin/thick-disc formation in 12 Milky-Way size simulations. Of these galaxies, four had significant mergers after star formation settles down, including two that are discussed above, \texttt{m12b}, and \texttt{m12f}.  These mergers do not destroy or disrupt the thin disc and contribute only in a second-order way to populating the thick disc.  The role of mergers in shaping activity during the bursty phase will be the topic of future work.

\subsection{Age Distributions}
\label{sec:sample_trend}

In this subsection, we briefly explore the age distributions for our spheroid, thick-disc, and thin-disc components in all 12 of the simulated galaxies listed in Table \ref{tab:one}.  Our aim is to illustrate how the ages of these components scale systematically with the bursty-phase lookback times for each galaxy.

\begin{figure}
    \centering 
	\includegraphics[width=0.98 \columnwidth, trim = 0.0 0.0 0.0 0.0]{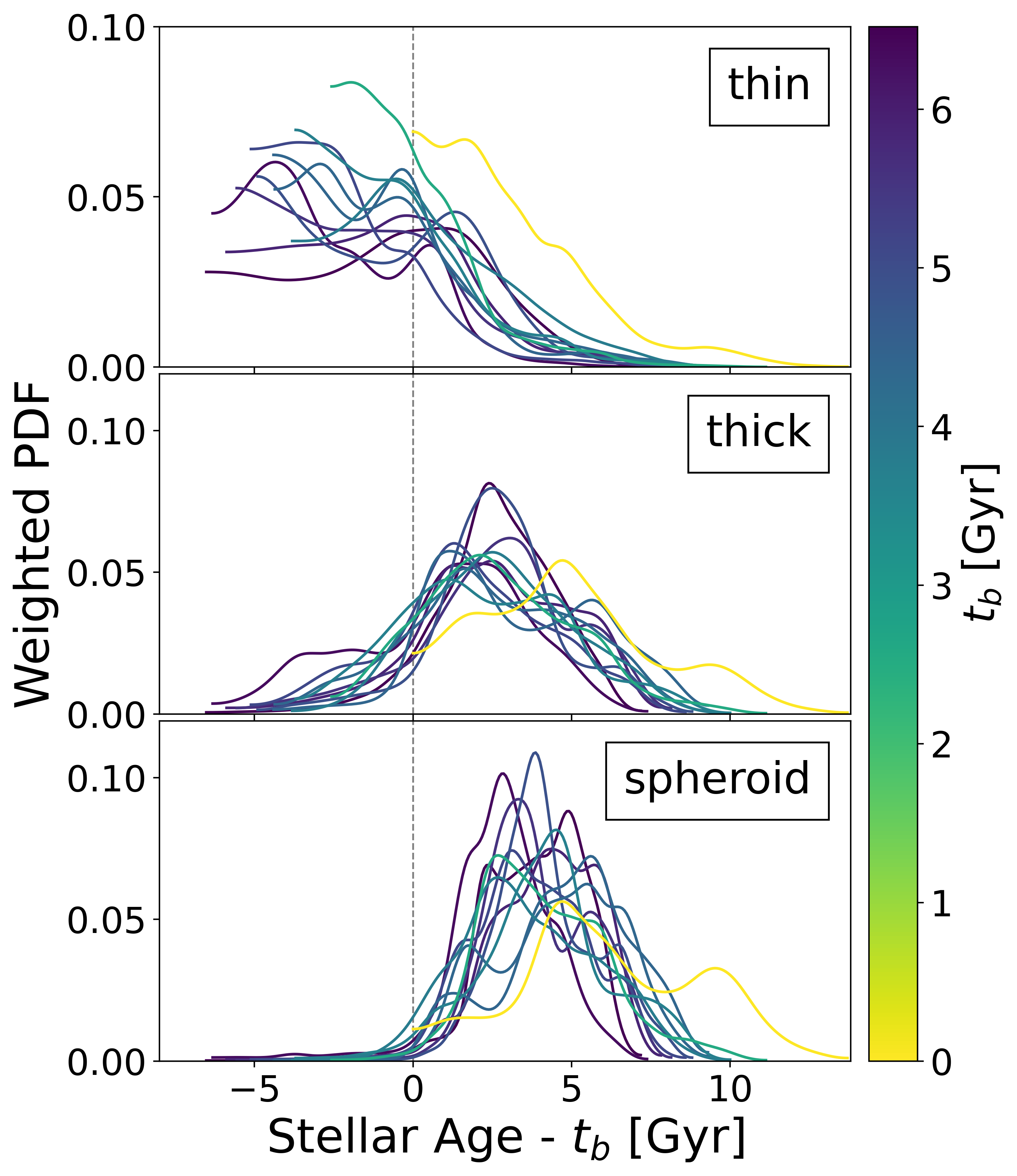}
    \caption[]{Age distributions of $z=0$ stars for 12 different Milky-Way-size galaxy simulations.  The ages are scaled relative to the bursty-phase lookback time $t_{\rm b}$ for each galaxy separately.  The top, middle and bottom panels show ages of star particles that are classified, based on orbital circularity, as thin-disc stars, thick-disc stars, and spheroid stars, respectively. Each galaxy has three lines, one in each panel, with a colour that maps to that galaxy's $t_{\rm b}$, as shown by the colour bar. When offset by $t_{\rm b}$, all galaxies show remarkable overlaps in the age distributions of all three populations.  The spheroid populations, in particular, have normalized age distributions that are quite similar.  Almost no {\it in situ} spheroid stars formed after $t_{\rm b}$. 
    }
	\label{fig:3comp_age_hist}
\end{figure}

Every line colour in the three panels of Figure \ref{fig:3comp_age_hist} corresponds to a different simulated galaxy.  The colour code is mapped to the bursty-phase lookback time of each galaxy.  These have values that range from $t_{\rm b} = 6.5$~Gyr (\texttt{Romeo}, purple) to $t_{\rm b} = 0$ (\texttt{m12w}, yellow).  Note that we have included \texttt{m12w} for completeness, even though it is still (barely) in its bursty phase by our definition. 

The top panel shows the stellar age distribution of thin-disc stars at $z=0$, classified by $\epsilon$, as illustrated in Figure \ref{fig:m12i_define_morph}. We have smoothed all lines with a Gaussian filter~\footnote{We use use kalepy.density \citep{Kelley2021} with bandwidth = $0.2$.}. The middle and bottom panels show thick-disc stars and spheroid stars, respectively. Importantly, all ages are plotted with respect to the bursty-phase lookback time. That is, we show stellar age - $t_{\rm b}$ instead of stellar age, in order to highlight the correspondence between the transition from bursty to steady star formation and the formation of different components in the galaxy.

When offset by $t_{\rm b}$, all galaxies and galaxy components show a remarkable similarity in their relative age distributions, especially the spheroid populations. Almost all the spheroid stars formed during the bursty phase (age - $t_{\rm b} >$  0), while thin-disc stars dominate after star formation has settled down (age - $t_{\rm b} <$ 0). As discussed above, one galaxy in our sample, \texttt{m12w}, never really settles down, so we have defined its busty phase to end at the present day ($t_{\rm b} = 0$ Gyr; yellow lines in Figure~\ref{fig:3comp_age_hist}). This likely explains why it is an apparent outlier.  This galaxy does have some stars with orbits that fall within our thin-disc category, but morphologically, its disc is much thicker than all of the other runs with a relatively small thin-disc fraction \citepalias{Yu2021}.

The thick-disc stars show an age distribution intermediate to thin-disc and spheroid stars. The distributions all peak prior to the end of the bursty phase but slightly closer to the end of the bursty phase than do the spheroid age distributions.  Though the distributions are similar, thick-disc stars have a tendency to be slightly younger than spheroid stars in most galaxies.  In addition, while the majority of thick-disc stars form during the bursty phase, there is a tail of slightly younger thick-disc stars that formed after the bursty phase ended.  This is not seen in the spheroid population.  Based on the analysis presented above, the tail of younger, post-bursty-phase, thick-disc stars is likely populated by stars that were born thin and subsequently heated to have tilted and more elongated orbits. 

The above discussion is consistent with the broad picture suggested by the analysis in previous sections: the orbital properties of $z=0$ stars in Milky-Way-size galaxies track an age sequence linked closely to the transition between bursty and steady star formation in galactic evolution.   Isotropic spheroid populations in Milky-Way-mass galaxies are formed early, from stars that are born on fairly radial orbits when star formation was irregular and bursty. Thin-disc populations form late, after the bursty phase has ended. Thick-disc stars are an intermediate age population, the bulk of which form prior to the end of the bursty phase, but slightly later than the isotropic spheroid.  If these simulations are correct, then spheroid age and thick-disc age should be tightly correlated with the bursty-phase lookback time. 

\section{Summary of Results}
\label{sec:conclusions}

We have used FIRE-2, Milky-Way-mass galaxy simulations to track the orbital evolution of stars from birth to the present day. 
As illustrated in Figure \ref{fig:m12i_define_morph}, we use stellar orbital properties at late times to classify stars with  \textit{thin-disc}, \textit{thick-disc}, and \textit{isotropic spheroid}.

All of our galaxies have an early, bursty star formation period that transitions to a steady, more uniform star formation rate at late times (Figures \ref{fig:m12i_define_param} and \ref{fig:sfh_jjc_CGM_evolve_all}). During the earliest period of the bursty phase, galaxies have clumpy, irregular morphologies (Figure \ref{fig:Romeo_Juliet_jjc_morph_evolve}) and stars are born on fairly radial orbits.  These stars evolve to inhabit an isotropic spheroid population today (Figures \ref{fig:Romeo_Juliet_jjc_morph_evolve} and  \ref{fig:m12i_jzjc_evolve}).  

At late times, after galaxies transition from bursty to steady star formation, new stars are born on extremely circular orbits along a narrow plane and the galaxy begins to build a substantial thin disc (Figure \ref{fig:m12i_jzjc_evolve}).  Prior to the transition from bursty to steady star formation, new stars begin to show more substantial rotation than in the bursty phase. During this ``spin-up'' phase, young stars have more circular orbits in the median, though there is a fair amount of scatter and orbital misalignment.  An example illustrating this is shown Figure \ref{fig:m12i_jzjc_evolve} at a lookback time of $\sim 7$ Gyr. Stars formed during this phase contribute substantially to thick-disc components at the present day.   We have explicitly use the term ``spin-up" to make connection with the observational findings of \cite{BK22}, who used it in connection with a transition epoch they identified in Milky Way stars at metallicity  $[{\rm Fe/H}] \simeq -1$.  Below this metallicty, they find that Milk Way stars have median tangential velocity of $\sim 0$ \kms\ (typical of a spheroid).  At $[{\rm Fe/H}] \simeq -1$, stars quickly transition to median tangential velocities of $\sim 100$ \kms\, which his typical of a {\em thick} disc.

In summary, and as illustrated in Figure \ref{fig:Romeo_Juliet_jjc_morph_evolve}, our galaxies progress through three stages in cosmic evolution:  1) a very early, bursty-irregular, chaotic phase; 2) a later, bursty-disc, ``spin-up" phase; and 3) a late-time, thin-disc, ``cool-down"  phase. Stars born during the first two stages evolve into isotropic spheroid and thick-disc populations today.  One corollary to this time sequence is that our thick-disc and spheroid populations are mostly formed at early times prior to the formation of the thin disc. As detailed in Figures \ref{fig:tracking_circ_full} - \ref{fig:tracking_theta_full}, stellar populations have median orbital properties at $z=0$ that are quite similar to their median orbital properties at birth. 

While birth orbit appears to be the most important driver in predicting where stars end up at $z=0$, mergers and secular heating {\em do} affect the dynamics of stars at late times in our simulations. Figures \ref{fig:delta_perGyr}, \ref{fig:change_in_delta1}, and \ref{fig:change_in_delta2} explore the evolutionary ``thickening'' after birth. Most mono-age populations experience the most orbital heating/torquing right after they form. After $\sim 0.5$ Gyr, the rate of heating drops and the median orbits become fairly stable.  The majority of -- though not all -- stars born with thin-disc orbits remain in thin-disc orbits at the present day.  There is some degree of heating that occurs after the steady/thin-disc phase begins, and these heated stars populate the youngest-age distribution tail of the thick disc.

Of course, the scenario we discuss here, especially as it relates to thin or thick-disc orbital evolution, would change in the event of major late-time mergers.  However, for galaxies that are Milky-Way size and smaller, we expect that the major merger rate will be quite rare at late times \citep{Stewart09,FMB10,RG16,Husko22}. While our sample of galaxies is small, there was no specific selection for systems without major mergers.  Half of the 12 galaxies in our sample experience mergers with stellar mass ratios greater than $0.15$ since $z=3$. All of these large mergers occur before $z=1$.  None of our 12 galaxies have a merger with mass ratio larger than $0.15$ since the onset of the thin-disc phase.
These statistics are consistent with the expectations from larger samples. For example, \citet{RG16} used used a cosmological sample of haloes to
show that that the major merger rate  (involving stellar mass ratios $>0.25$) for Milky-Way size galaxies at late times is only $\sim 0.02$ Gyr$^{-1}$.  This suggests that, while rare, significant late-time mergers should occasionally occur at this mass scale. In those rarer cases, especially those involving close to one-to-one mass ratio mergers, we would expect orbits to be randomized and destroy discs. 

Two of the five primary simulations we have analyzed experience substantial, though still minor (LMC-size). late-time mergers during the steady phase (see the arrows in right panels of Figure \ref{fig:tracking_circ_full}). These mergers, with stellar mass ratios $\sim 0.1$, neither destroy thin discs nor significantly alter the orbital properties of young stars. Mergers do appear to heat some thin-disc stars enough to populate thick-disc components, but only by populating the tail of youngest thick-disc stars. An analysis like ours performed on a much larger sample of galaxies that include significantly larger later-time mergers will be required to understand how and how often late-time major mergers affect the picture presented here.

As discussed in the introduction, there have been a series of FIRE-2 papers that have examined the relationship between star formation burstiness, galaxy kinematics, galaxy morphology, and  the development of a hot gaseous halos around galaxies \citep{Ma2017,Stern20,Yu2021,Gurvich2022,Hafen2022}. In particular, all galaxies of sufficiently high mass in these simulations experience a fairly sharp transition from bursty  to steady star formation.  The lookback time to this transition, $t_{\rm B}$, is different for each galaxy, coincides with the virialization of the inner CGM \citep{Stern20,Yu2021}, and also correlates with distinct changes in the angular momentum properties of both the CGM and ISM \citep{Hafen2022,Gurvich2022}.

The lookback time to the bursty-to-steady transition, $t_{\rm B}$, sets a characteristic age for morphological components in our galaxies at $z=0$ (Figure \ref{fig:3comp_age_hist}).  Specifically, virtually all {\em in situ} isotropic spheroid stars form prior to $t_{\rm B}$.  Moreover, when stellar ages are shifted with respect to $t_{\rm B}$ for each galaxy, their age distributions are remarkably similar. This means, for example, that the youngest stars in the {\it in situ} spheroid should have ages that map to the bursty-to-steady transition time, which corresponds to the time of inner CGM virialization.   As shown  in Figure \ref{fig:3comp_age_hist} and \citet{Yu2021}, the median ages of thick-disc stars are also tightly correlated with $t_{\rm B}$.  The youngest tail of heated thick-disc stars are formed after $t_{\rm B}$.  The majority of thin-disc stars form after $t_{\rm B}$.  These trends provide a potential avenue for linking the observed age distributions of stars in the Milky Way and similar local galaxies to the nature of {\it in situ} star formation over cosmic time.

Future work that compares predicted distributions of galaxy morphologies and kinematic properties to observed characteristics at intermediate and high redshft \citep{Smit18, Zhang19, Wisnioski19, Wu22, Ferreira22, Robertson22} will provide useful tests of these simulations.

\section*{Acknowledgements}
SY and JSB were supported by NSF grants AST-1910346 and AST-1518291. 
JS was supported by the Israel Science Foundation (grant No. 2584/21). MBK acknowledges support from NSF CAREER award AST-1752913, NSF grants AST-1910346 and AST-2108962, NASA grant 80NSSC22K0827, and HST-AR-15809, HST-GO-15658, HST-GO-15901, HST-GO-15902, HST-AR-16159, and HST-GO-16226 from the Space Telescope Science Institute, which is operated by AURA, Inc., under NASA contract NAS5-26555. 
CAFG was supported by NSF through grants AST-1715216, AST-2108230,  and CAREER award AST-1652522; by NASA through grants 17-ATP17-006 7 and 21-ATP21-0036; by STScI through grants HST-AR-16124.001-A and HST-GO-16730.016-A; by CXO through grant TM2-23005X; and by the Research Corporation for Science Advancement through a Cottrell Scholar Award.
AW received support from: NSF via CAREER award AST-2045928 and grant AST-2107772; NASA ATP grant 80NSSC20K0513; HST grants AR-15809, GO-15902, GO-16273 from STScI.
We ran simulations using: XSEDE, supported by NSF grant ACI-1548562; Blue Waters, supported by the NSF; Pleiades, via the NASA HEC program through the NAS Division at Ames Research Center. 
Allocations AST21010 and AST20016 were supported by the NSF and TACC.  

We thank an anonymous referee for several suggestions that helped our presentation significantly.  We also acknowledge Charlie Conroy, David Weinberg, Andrey Kravtsov, Vasily Belokurov, Ana Bonaca for inspiring discussions.  Patrick Staudt, Stefani Germanotta, and Jeppe  Laursen suggested we were on the right track during initial stages of this research. 

\section*{Data Availability}
The data supporting the plots within this article are available on reasonable request to the corresponding author. A public version of the {\small GIZMO} code is available at \url{http://www.tapir.caltech.edu/~phopkins/Site/GIZMO.html}. FIRE-2 simulations are publicly available \citep{Wetzel2022} at \url{http://flathub.flatironinstitute.org/fire}. Additional data including simulation snapshots, initial conditions, and derived data products are available at \url{https://fire.northwestern.edu/data/}. Some of the publicly available software packages used to analyze these data are available at: \url{https://bitbucket.org/awetzel/gizmo\_analysis}, and \url{https://bitbucket.org/awetzel/utilities}.




\bibliographystyle{mnras}
\bibliography{ref} 




\appendix

\section{Co-evolution in star formation, young-star circularity, and the CGM}
\label{app:1}
\begin{figure*}
    \centering
	\includegraphics[width=0.45 \textwidth, trim = 0.0 0.0 0.0 0.0]{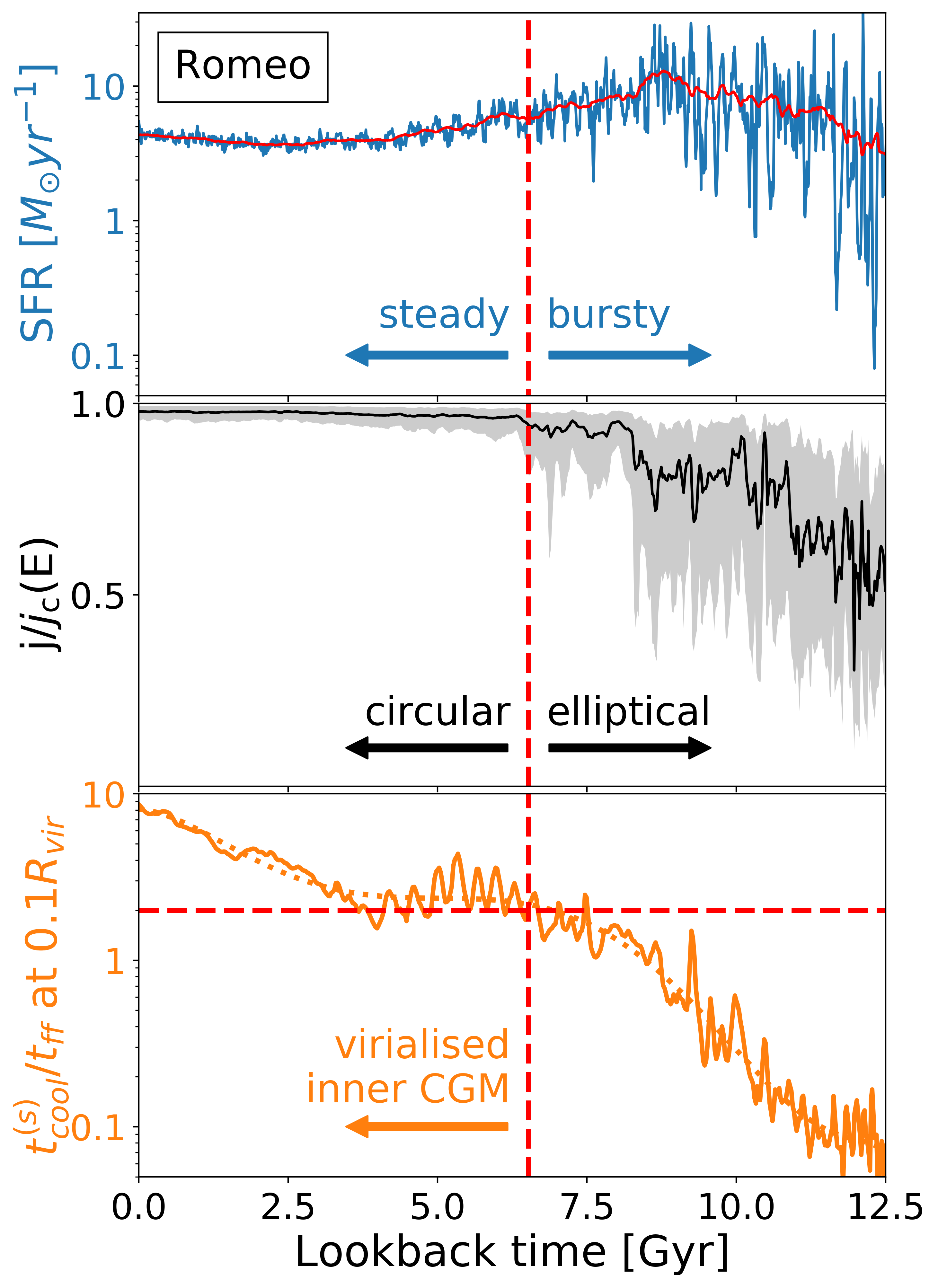}
	\includegraphics[width=0.45 \textwidth, trim = 0.0 0.0 0.0 0.0]{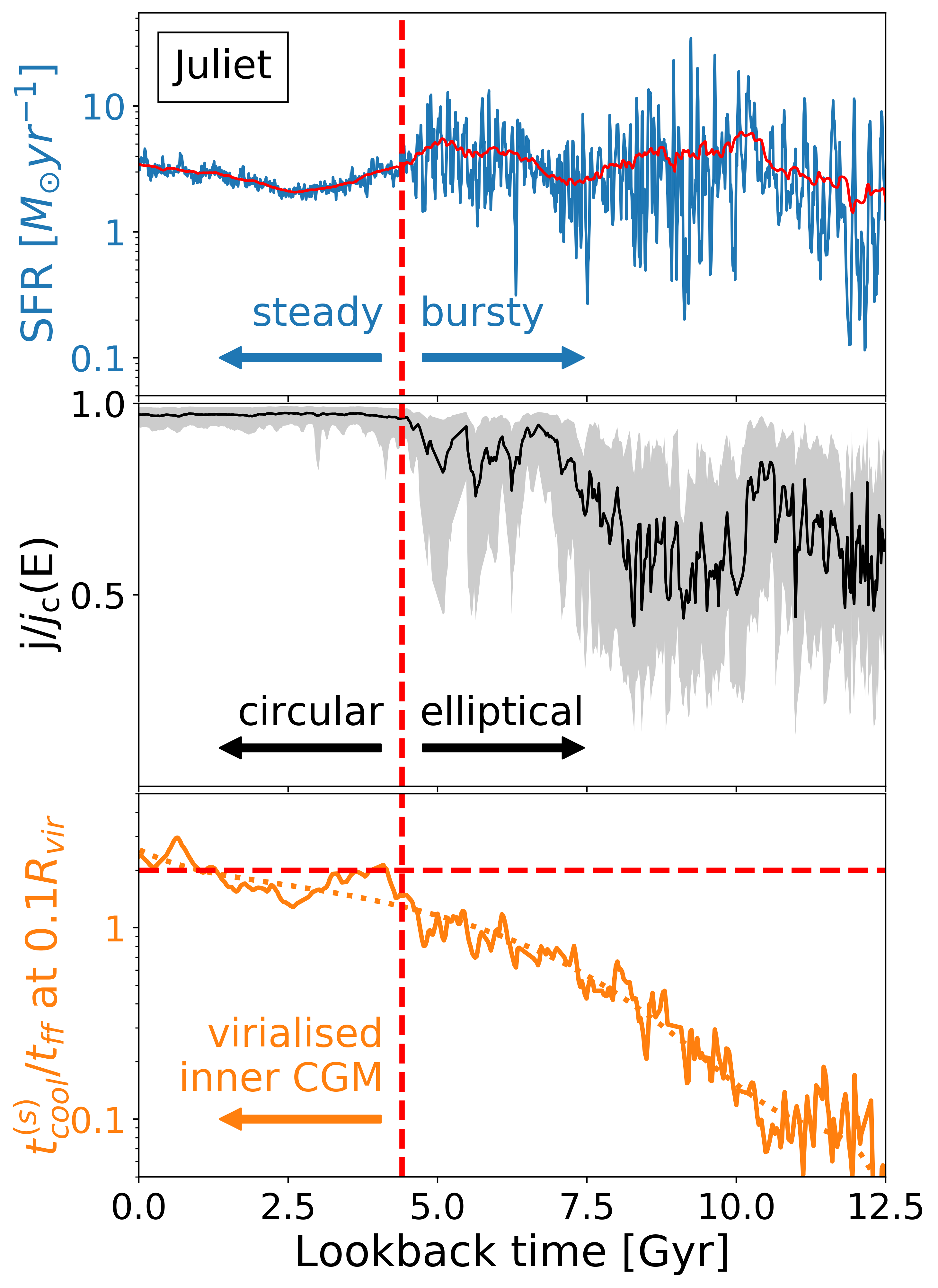}
	\includegraphics[width=0.45 \textwidth, trim = 0.0 0.0 0.0 0.0]{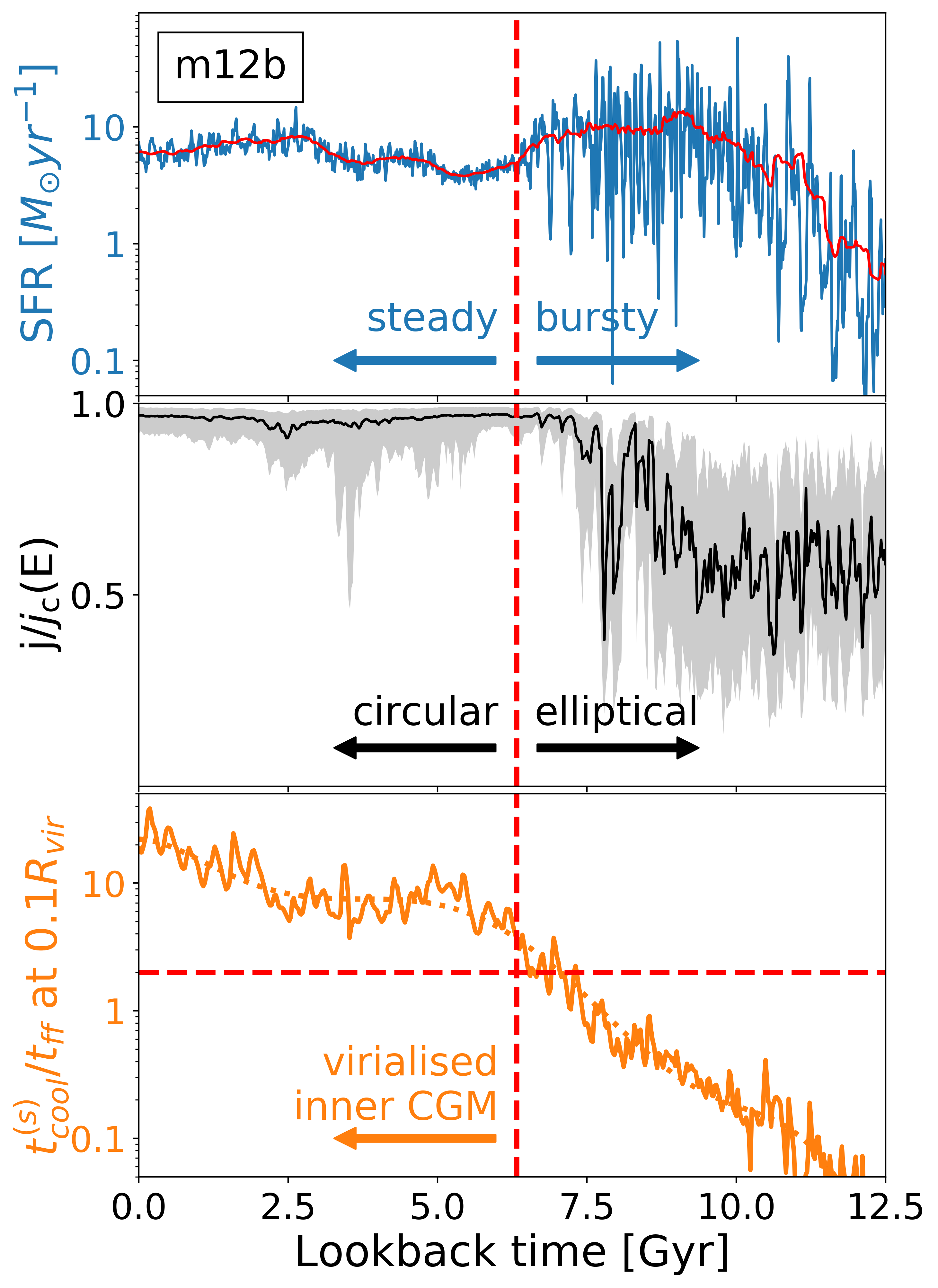}
	\includegraphics[width=0.45 \textwidth, trim = 0.0 0.0 0.0 0.0]{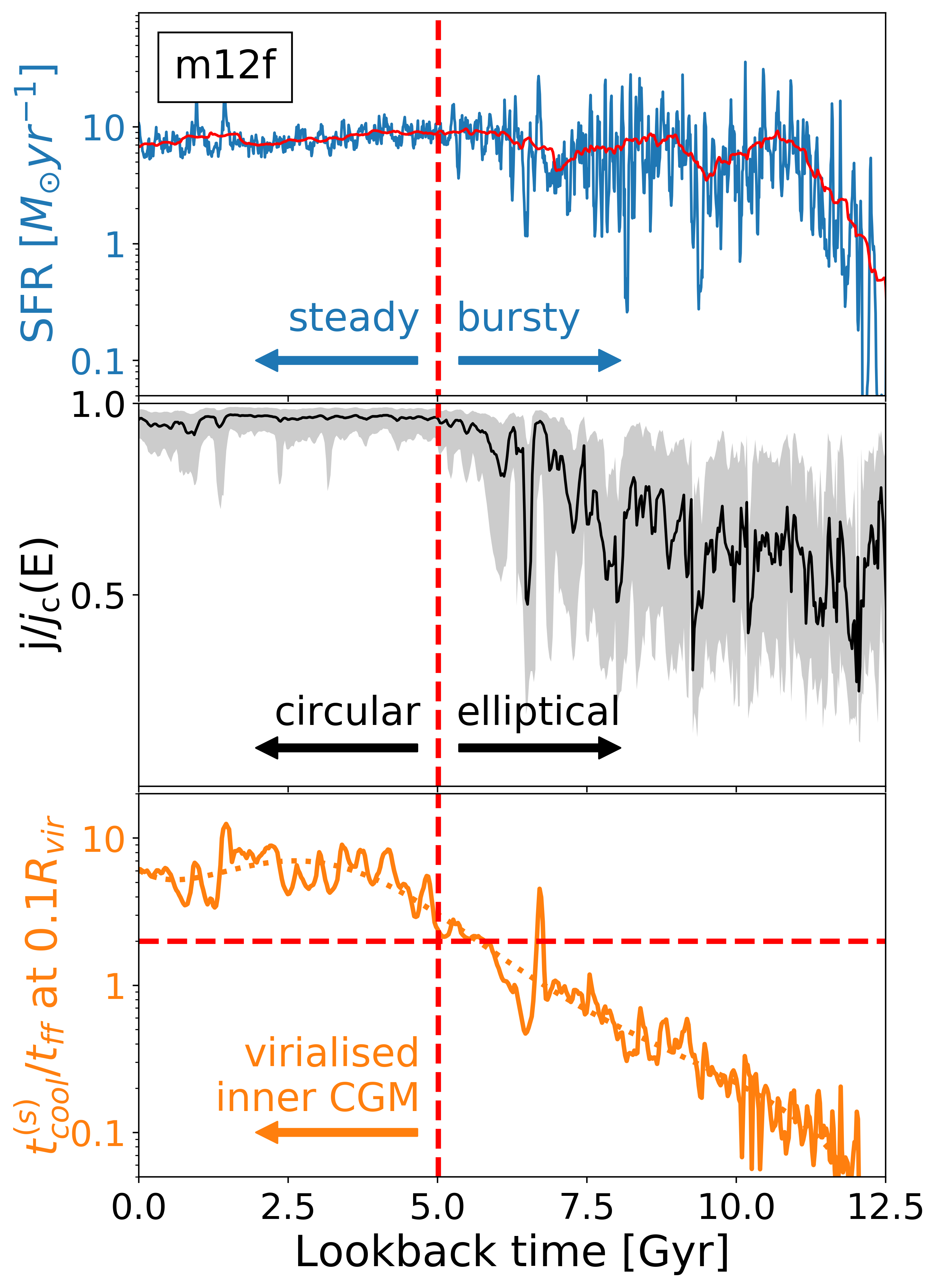}
    \caption[]{The top portions of each panel show the star formation rate as a function of lookback time.  The middle portions show the orbital circulariites of young stars. The bottom portions show the CGM virialisation propensity as functions of lookback time.  The four panels show results for \texttt{Romeo}, \texttt{Juliet}, \texttt{m12b}, and \texttt{m12f}. This figure is similar to Figure \ref{fig:m12i_define_param} in the main text.
    }
	\label{fig:sfh_jjc_CGM_evolve_all}
\end{figure*}

Each panel of Figure \ref{fig:sfh_jjc_CGM_evolve_all} is similar to
Figure \ref{fig:m12i_define_param} in the main text, but shows results for the other four galaxies in our primary sample.  The top panels track the star formation rate as a function of lookback time, with a clear transition from bursty to steady star formation marked by the vertical dashed red line (at $t_{\rm B}$).   The middle panel shows the median (solid) and one-sigma distribution (grey shaded band) of 3D circularities for all young stars (age < 100 Myr) as a function of lookback time.  The third panel shows the evolution of the ratio $\tcoolsh / \tff $, which tracks the propensity of the inner CGM to be virialized. The ratio  $\tcoolsh / \tff = 2$ is marked by a horizontal dashed line.  During the early, bursty phase, we see that $\tcoolsh/\tff \ll 1$.  When this condition is met, the inner CGM is dominated by the supersonic infall of cold, often clumpy, gas.  At late times, $\tcoolsh/\tff \gtrsim 2$, and the GCM becomes hot, smooth, and largely supported by thermal pressure.

During the steady phase, stars are born on quite circular orbits, very close to $\epsilon_{3D} = 1$, with very little scatter.  During the early bursty phase, the scatter is much higher.  A few billion years before the bursty phase ends, we see a gradual ``spin up'' in the orbits of young stars, as the circularities rise towards unity.  Note that \texttt{m12b} and \texttt{m12f} (bottom two panels)  Both  undergo $\sim 1/10$ merger events during the steady phase, at lookback times of $\sim 3$ Gyr and $\sim 0.3$ Gyr, respectively.  At these times, the distribution of young stellar orbits in both galaxies becomes slightly more radial,  but the period is fairly brief.  The feature is more pronounced  in \texttt{m12b}, where the merging orbit is polar.  The merger in \texttt{m12f} is prograde.

\section{Heating Rates}
\label{app:2}

In subsection \ref{sub:evolve}, we discussed the evolution of the orbital properties of mono-age stellar  populations with cosmic time, and presented the average rate of change in orbital properties from birth to the present day in Figure \ref{fig:delta_perGyr}.  We specifically explored the change in circularity $\epsilon$, 3D circularity $\epsilon_{\rm 3D}$,  and alignment angle $\theta$, per Gyr.

\begin{figure*}
    \centering
	\includegraphics[width=0.45 \textwidth, trim = 0.0 0.0 0.0 0.0]{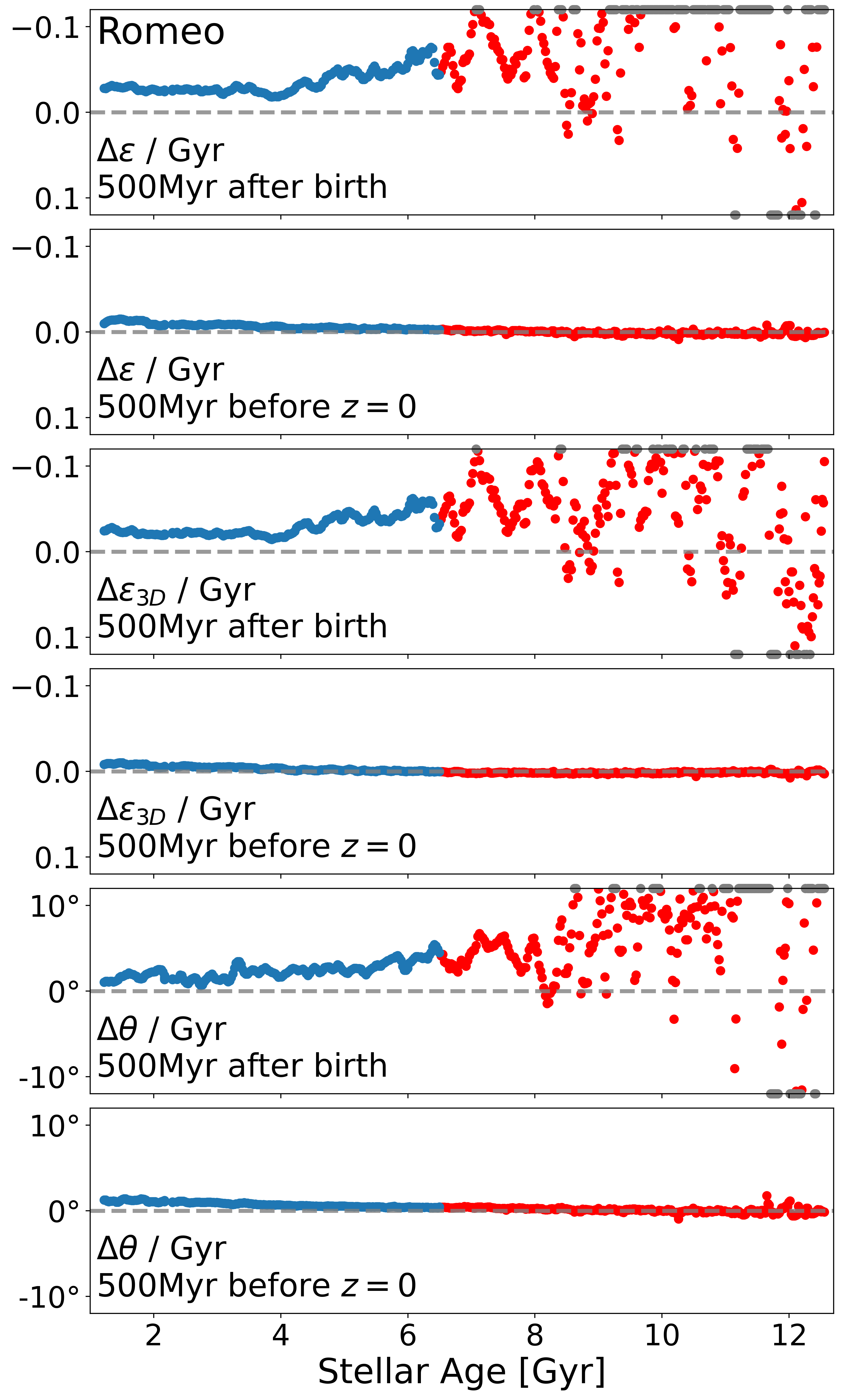}
	\includegraphics[width=0.45 \textwidth, trim = 0.0 0.0 0.0 0.0]{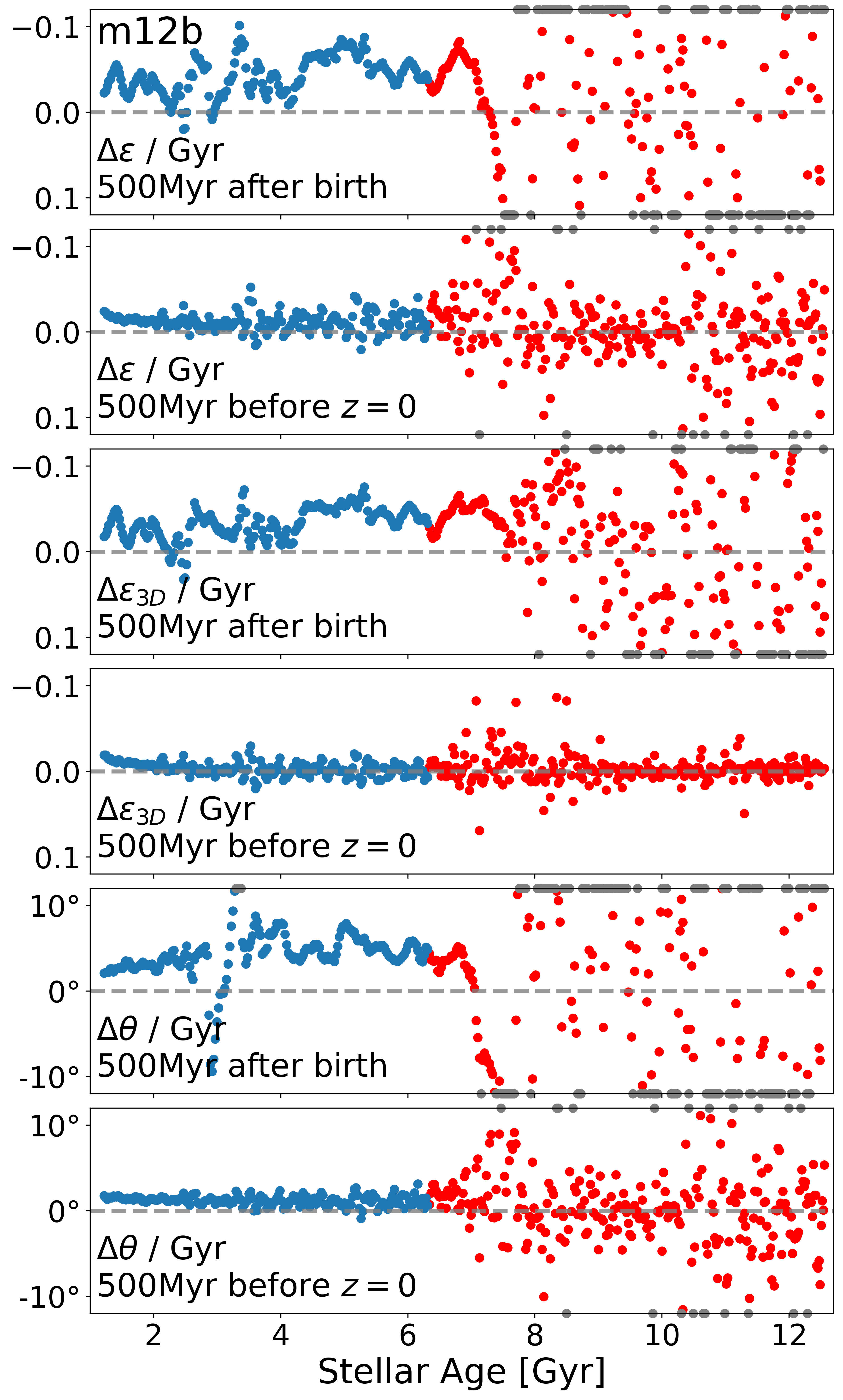}
    \caption[ ``Instantaneous'' thickening/heating/torquing effect for different stellar populations born at different times in\texttt{Romeo} and \texttt{m12b} ]
    {
    ``Instantaneous'' thickening/heating/torquing effect for different stellar populations born at different times in \texttt{Romeo} (left) and \texttt{m12b} (right). Similar to Figure \ref{fig:delta_perGyr}, here we quantify the amount of thickening/heating/torquing happening for different populations at birth and now by calculating the change in circularity $\epsilon$, 3D circularity $\epsilon_{\rm 3D}$, and alignment angle $\theta$ 500 Myr after alignment angle r birth and 500 Myr before $z=0$. For all three quantities, the change happens mostly right after birth while for the period of time before $z=0$, the change for all populations, especially the ones born in steady phase, is almost zero. This could also be seen in Figure \ref{fig:delta_perGyr} as the lines are getting flatter around 0 Gyr.
    }
	\label{fig:change_in_delta1}
\end{figure*}

\begin{figure*}
    \centering
	\includegraphics[width=0.45 \textwidth, trim = 0.0 0.0 0.0 0.0]{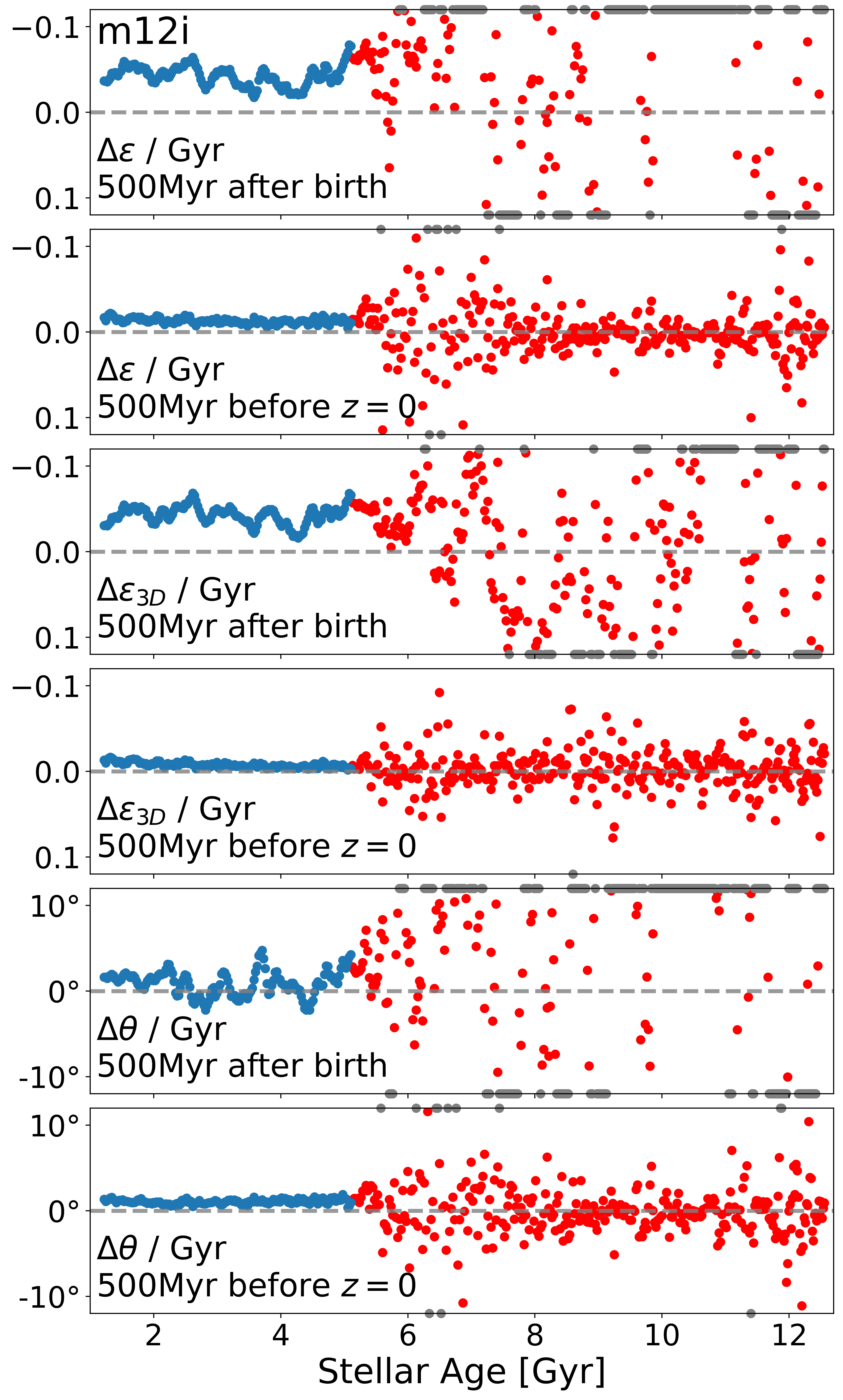}
	\includegraphics[width=0.45 \textwidth, trim = 0.0 0.0 0.0 0.0]{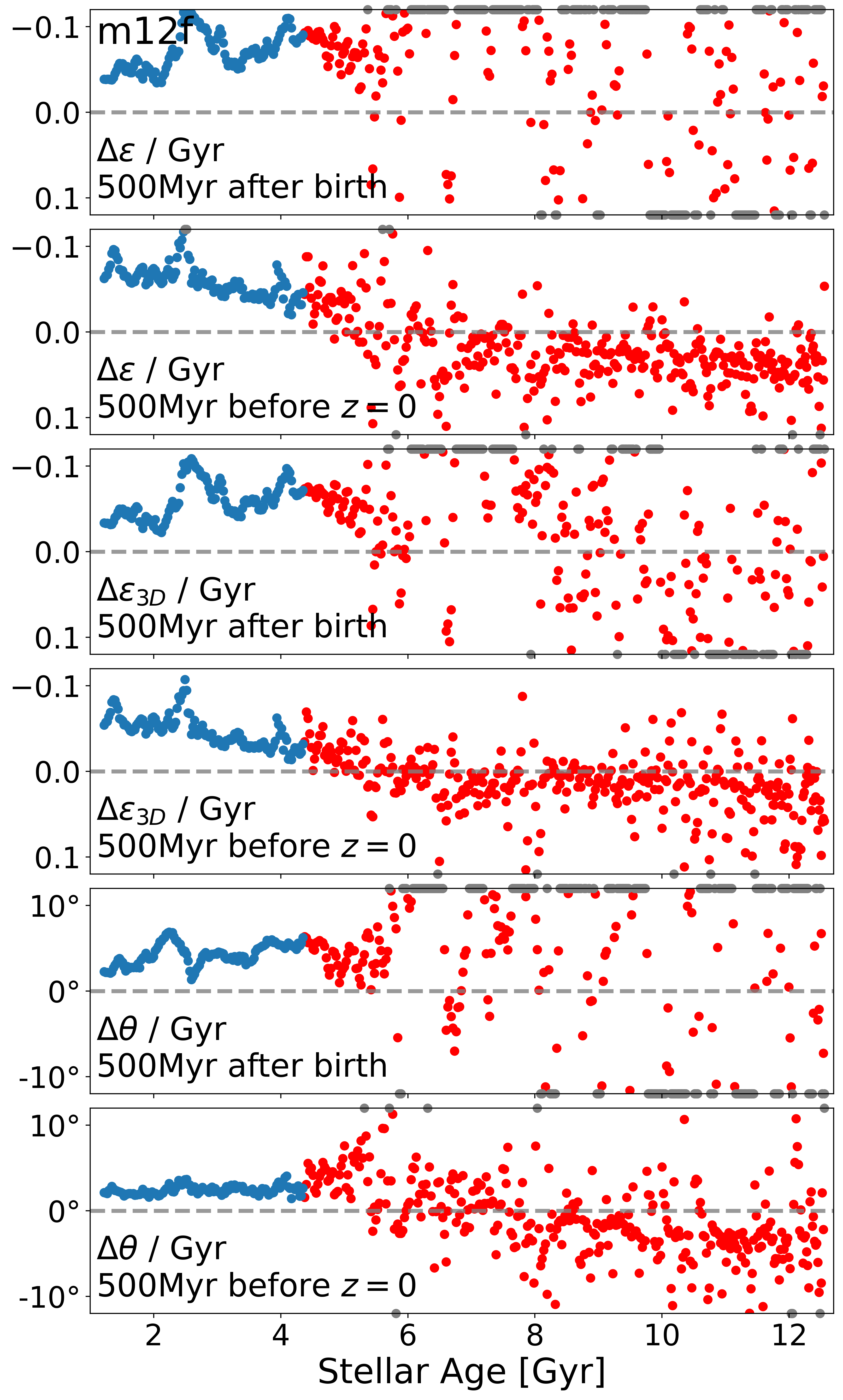}
    \caption[ ``Instantaneous'' thickening/heating/torquing effect for different stellar populations born at different times in\texttt{m12i} and \texttt{m12f} ]
    {
    Same as Figure \ref{fig:change_in_delta1}, now for \texttt{m12i} (left) and \texttt{m12f} (right). Note that, for \texttt{m12f}, the change around $z=0$ is still large and this is due to the fact that m12f has recently undergone a merger (as indicated by the arrow in Figure \ref{fig:tracking_circ_full}, Figure \ref{fig:tracking_circ3D_full}, and Figure \ref{fig:tracking_theta_full}). The recent merger would perturb the orientation and total mass of the central galaxy, resulting in fluctuation in all three quantities.
    }
	\label{fig:change_in_delta2}
\end{figure*}

Here, in Figures \ref{fig:change_in_delta1} and  \ref{fig:change_in_delta2}, we present the rate of change measured over shorter timescale to understand how the heating rates evolve over time.  Instead of averaging over the entire age of a stellar population, we select 500-Myr time windows right after stars form and right before $z=0$. Since the time period we choose is relatively small compared to the cosmological timescale of the transition in the galaxy evolution, it is reasonable to think of these two values as the ``instantaneous'' thickening/heating/torquing effect. 

Red dots represent stellar populations born during the bursty phase and blue dots show for stars that form after star formation settles down. The grey dashed lines align with zero, to guide the eye.  Dots that lie on this line have no change in median $\epsilon$, $\epsilon_{\rm 3D}$, or $\theta$ during the time period explored.

Generally speaking, the instantaneous rates are lowest near $z=0$ and higher just after birth.  Stars born during the bursty phase (red dots) have significantly higher rates of orbital evolution just after birth. 

Together with the results from Section~\ref{sec:circ_evolve}, these results suggests that disc heating/torquing is not significant enough to turn most thin-disc stars into thick-disc stars and thus is unlikely to be the primary formation mechanism for the bulk of the thick-disc stars in the Milky-Way-mass systems.

\bsp	
\label{lastpage}
\end{document}